\documentclass[pre,aps,reprint,twocolumn,amsmath]{revtex4}

\usepackage{graphicx}
\usepackage{psfrag}

\begin{document}

\title{Social cycling and conditional responses in the Rock-Paper-Scissors 
  game\footnote{ZW and BX contributed equally to this work.
    ZW, BX designed and performed experiment;
    HJZ, BX constructed theoretical model;
    HJZ developed analytical and numerical methods;
    BX, ZW, HJZ analyzed and interpreted data;
    HJZ, BX wrote the paper.
    Correspondence should be addressed to HJZ ({\tt zhouhj@itp.ac.cn}).}
}

\author{Zhijian Wang$^{1}$, Bin Xu$^{2,1}$, and Hai-Jun Zhou$^{3}$}
\affiliation{
$^{1}$Experimental Social Science Laboratory,
    Zhejiang University, Hangzhou 310058, China \\
$^{2}$Public Administration College,
    Zhejiang Gongshang University, Hangzhou 310018, China\\
$^{3}$State Key Laboratory of Theoretical Physics,
    Institute of Theoretical Physics, Chinese Academy of Sciences, 
    Beijing 100190, China
}
  
\begin{abstract}
  How humans make decisions in non-cooperative strategic interactions is a
  challenging question. For the fundamental model system of Rock-Paper-Scissors (RPS)
  game, classic game theory of infinite rationality predicts the Nash equilibrium (NE)
  state with every player randomizing her choices to avoid being exploited, while
  evolutionary game theory of bounded rationality in general
  predicts persistent cyclic motions, especially for finite populations. However, as
  empirical studies on human subjects have been relatively sparse, it is still a
  controversial issue as to which theoretical framework is more appropriate to
  describe decision making of human subjects. Here we observe population-level
  cyclic motions in a laboratory experiment of the discrete-time iterated RPS game
  under the traditional random pairwise-matching protocol. The cycling direction and
  frequency are not sensitive to the payoff parameter a. This collective behavior
  contradicts with the NE theory but it is quantitatively explained by a microscopic
  model of win-lose-tie conditional response without any adjustable parameter. Our
  theoretical calculations reveal that this new strategy may offer higher payoffs to
  individual players in comparison with the NE mixed strategy, suggesting that high
  social efficiency is achievable through optimized conditional response. 
\end{abstract}

\maketitle

The Rock-Paper-Scissors (RPS) game is a fundamental non-cooperative
game. It has been widely used to study competition phenomena in society and
biology, such as species diversity of ecosystems
\cite{Sinervo-Lively-1996,Kerr-etal-2002,Semmann-Krambeck-Milinski-2003,Lee-etal-2005,Reichenbach-Mobilia-Frey-2007,Allesina-Levine-2011}
and price dispersion of markets
\cite{Maskin-Tirole-1988,Cason-Friedman-2003}.
This game has three candidate actions $R$ (rock), $P$ (paper) and $S$
(scissors). In the simplest settings the payoff matrix is characterized by a single 
parameter, the payoff $a$ of the winning action ($a>1$, see Fig.~1A)
\cite{Hoffman-etal-2012}. There are
the following non-transitive dominance relations among the actions:
$R$ wins over $S$, $P$ wins over $R$, yet $S$ wins over $P$ (Fig.~1B). Therefore
no action is absolutely better than the others.

The RPS game is also a basic model system for studying decision making
of human subjects in competition environments. Assuming ideal rationality
for players who repeatedly playing the RPS game within a population, 
classical game theory predicts that individual players will completely
randomize their action choices so that their behaviors will be unpredictable and
not be exploited by the other players
\cite{Nash-1950,Osborne-Rubinstein-1994}.
This is referred to as the mixed-strategy Nash equilibrium
(NE), in which  every player chooses the three actions with equal
probability $1/3$ at each game round. When the payoff parameter $a<2$ this
NE is evolutionarily unstable with respect to small perturbations but it becomes
evolutionarily stable at $a>2$ 
\cite{Taylor-Jonker-1978}.
On the other hand, evolutionary game theory drops the infinite rationality assumption
and looks at the RPS game from the angle of
evolution and adaption
\cite{MaynardSmith-Price-1973,MaynardSmith-1982,Nowak-Sigmund-2004,Sandholm-2010}.
Evolutionary models based on various microscopic
learning rules (such as the replicator dynamics 
\cite{Taylor-Jonker-1978},
the best response dynamics
\cite{Matsui-1992,Hopkins-1999}
and the logit dynamics
\cite{Blume-1993,Hommes-Ochea-2012})
generally predict cyclic evolution patterns for the action marginal distribution
(mixed strategy) of each player, especially for finite populations.

Empirical verification of non-equilibrial persistent cycling in the human-subject
RPS game (and other non-cooperative games) has been rather nontrivial, as the
recorded evolutionary trajectories are usually highly stochastic and not long
enough to draw convincing conclusions.
Two of the present authors partially overcame these difficulties 
by using social state velocity vectors \cite{Xu-Wang-2011} and 
forward and backword transition vectors \cite{Xu-Wang-2012}
to visualize violation of detailed balance in game evolution trajectories.
The cycling frequency of directional flows in the neutral RPS game ($a=2$) was
later quantitatively measured in \cite{Xu-Zhou-Wang-2013} using a coarse-grained
counting technique.
Using a cycle rotation index as the order parameter, 
Cason and co-workers \cite{Cason-Friedman-Hopkins-2014} also obtained
evidence of persistent cycling in some evolutionarily stable RPS-like games, 
if players were allowed to update actions  asynchronously in continuous time and 
were informed about the social states of
the whole population by some sophisticated `heat maps'.

In this work we investigate whether cycling is a general aspect of the
simplest RPS game.  We adopt an improved cycle counting method on the
basis of our earlier experiences \cite{Xu-Zhou-Wang-2013} and study directional
flows in evolutionarily stable ($a>2$) and unstable ($a<2$) discrete-time RPS games.
We show strong evidence that the RPS game is an 
intrinsic non-equilibrium system, which cannot
be fully described by the NE concept even in the evolutionarily
stable region but rather exhibits persistent population-level
cyclic motions. 
We then bridge the collective cycling behavior and the highly stochastic 
decision-making of individuals through a simple
conditional response (CR) mechanism. Our empirical
data confirm the plausibility of this microscopic model of
bounded rationality.
We also find that if the transition parameters of the CR strategy are
chosen in an optimized way, this strategy will outperform the NE
mixed strategy in terms of the accumulated payoffs of
individual players, yet the action marginal
distribution of individual players is indistinguishable from that
of the NE mixed strategy. We hope this work will stimulate
further experimental and theoretical studies on the microscopic
mechanisms of decision making and learning in basic game systems and
the associated non-equilibrium behaviors
\cite{Castellano-Fortunato-Loreto-2009,Huang-2013}.

\section*{Results}

\subsection*{Experimental system}
We recruited a total number of $360$ students from different
disciplines of Zhejiang University to form
$60$ disjoint populations of size $N=6$.
Each population then carries out one experimental session by
playing the RPS game $300$ rounds (taking $90$--$150$ minutes)
with a fixed value of $a$.
In real-world situations, individuals often have to
make decisions
based only on partial input information.
We mimic such situations in our laboratory experiment
by adopting the traditional random pairwise-matching
protocol \cite{Osborne-Rubinstein-1994}:
At each game round (time) $t$ the players are randomly paired within
the population and compete  with their pair opponent once;
after that each player gets feedback information about her
own payoff as well as her and her
opponent's action.
As the whole experimental session finishes, the players are paid
in real cash proportional to their accumulated payoffs
(see {\em Materials and Methods}).
Our experimental
setting differs from those of two other recent experiments,
in which every player competes against the whole
population \cite{Hoffman-etal-2012,Cason-Friedman-Hopkins-2014}
and may change actions in continuous time \cite{Cason-Friedman-Hopkins-2014}.
We set $a=1.1$, $2$, $4$, $9$ and $100$, respectively,
in one-fifth of the populations so as to compare
the dynamical behaviors in the evolutionarily unstable,
neutral, stable and deeply stable regions.

\subsection*{Action marginal distribution of individual players}
We observe that the individual players shift their actions frequently in
all the populations except one with $a=1.1$ (this
exceptional population is discarded from further analysis, see
Supporting Information).
Averaged among the $354$ players of these $59$ populations,
the probabilities that a player adopts action
$R$, $P$, $S$ at one game round are, respectively,
$0.36 \pm 0.08$, $0.33 \pm 0.07$ and $0.32 \pm 0.06$
(mean $\pm$ s.d.). We obtain very similar results
for each set of populations of the same $a$ value (see
Table~S1).
These results are consistent with NE
and suggest the NE mixed strategy is
a good description of a player's marginal distribution
of actions.
However, a player's actions at two consecutive
times are not independent but
correlated (Fig.~S1). At each
time the players are more likely to repeat their
last action than to shift action
either counter-clockwise
(i.e., $R\rightarrow P$, $P\rightarrow S$, $S\rightarrow R$,
see Fig.~1B)
or clockwise ($R\rightarrow S$, $S\rightarrow P$,
$P\rightarrow R$). This inertial effect is especially
strong at $a=1.1$ and it diminishes as $a$ increases.

\subsection*{Collective behaviors of the whole population}
The social state of the population
at any time $t$ is denoted as
${\bf s}(t) \equiv \bigl(n_R(t), n_P(t), n_S(t)\bigr)$
with $n_q$ being the number of players adopting action
$q \in \{R, P, S\}$.
Since $n_R+n_P+n_S \equiv N$ there
are $(N+1) (N+2)/2$ such social states, all lying
on a three-dimensional plane bounded by an
equilateral triangle  (Fig.~1C). 
Each population leaves a trajectory on this plane
as the RPS game proceeds.
To detect rotational flows,
we assign for every social state transition
${\bf s}(t) \rightarrow {\bf s}(t+1)$
a rotation angle $\theta(t)$, which measures the angle
this transition rotates with respect to the centroid
${\bf c}_0 \equiv (N/3, N/3, N/3)$ of the
social state plane \cite{Xu-Zhou-Wang-2013}.
Positive and negative $\theta$ values
signify counter-clockwise and clockwise rotations, respectively,
while $\theta = 0$ means the transition is not a rotation
around ${\bf c}_0$.
For example, we have $\theta (1) =\pi/3$, $\theta(2)=0$, 
and $\theta(3)=-2 \pi /3$ for the exemplar transitions shown in
Fig.~1C.

\begin{figure}
  \psfrag{n_s}{$n_S$}
  \psfrag{n_p}{$n_P$}
  \psfrag{n_r}{$n_R$}
  \psfrag{R}{{\small R}}
  \psfrag{P}{{\small P}}
  \psfrag{S}{{\small S}}
  \begin{center}
    \centerline{\includegraphics[width=0.4\textwidth]{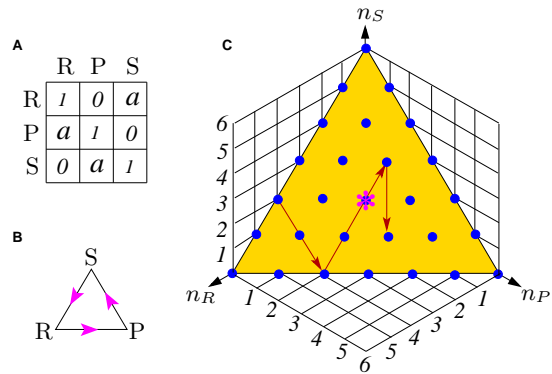}}
    \caption{
      The Rock-Paper-Scissors game.
      (A) Each matrix entry specifies the row action's payoff.
      (B) Non-transitive dominance relations
      ($R$ beats $S$, $P$ beats $R$, $S$ beats $P$) among the three actions.
      (C) The social state plane for a population of size $N=6$.
      Each filled circle denotes a social state $(n_R, n_P, n_S)$; 
      the star marks the centroid ${\bf c}_0$;
      the arrows indicate three social state transitions at 
      game rounds $t=1,2,3$.}
  \end{center}
\end{figure}

The net number of cycles around ${\bf c}_0$ during the time interval 
$[t_0, t_1]$ is computed by 
\begin{equation}
  C_{t_0,t_1} \equiv
  \sum_{t=t_0}^{t_{1}-1} \frac{ \theta(t)}{ 2 \pi }  \; .
\end{equation}
As shown in Fig.~2 (A-E),
$C_{1, t}$ has an increasing
trend in most of the $59$ populations, indicating
persistent counter-clockwise cycling.
The cycling frequency of each trajectory
in $[t_0, t_1]$ is evaluated by
\begin{equation}
  f_{t_0, t_1} \equiv \frac{ C_{t_0, t_1} }{ t_1-t_0}  \; .
\end{equation}
The values of $f_{1, 300}$ for all the $59$
populations are listed in Table~1, from
which we obtain the mean frequency
to be $0.031 \pm 0.006$ ($a=1.1$, mean $\pm$ SEM),
$0.027\pm 0.008$  ($a=2$),
$0.031 \pm 0.008$ ($a=4$),
$0.022 \pm 0.008$ ($a=9$) and
$0.018 \pm 0.007$ ($a=100$).
These mean frequencies are all positive irrespective to the
particular value of $a$, indicating that behind the seemingly
highly irregular social state evolution process, there is
a deterministic pattern of social state cycling from slightly rich in action $R$, to
slightly rich in  $P$, then to slightly rich in $S$, and then back to slightly
rich in $R$ again.
Statistical analysis
confirms that $f_{1, 300}>0$ is significant
for all the five sets of populations
(Wilcoxon signed-rank test, $p<0.05$).
The correlation between the mean cycling frequency $f_{1,300}$ and the
payoff parameter $a$ is not statistically significant 
(Spearman's correlation test: $r=-0.82$, $p=0.19$, for $n=5$ mean
frequencies; and 
$r=-0.16$, $p=0.24$, for $n=59$ frequencies).
We also notice that  the mean cycling frequency in the second half of the
game ($f_{151, 300}$) is slightly higher than
that in the first half ($f_{1,150}$) for all the five
sets of populations (Table~S2),
suggesting that cycling does not die out with time.

A recent experimental work \cite{Frey-Goldstone-2013}
also observed cycling behaviors in a RPS-like game with more than three actions.
Evidences of persistent cycling in some
complete-information and continuous-time RPS-like games were
 reported in another experimental study
\cite{Cason-Friedman-Hopkins-2014}.
However, no (or only very weak) evidence of population-level cycling was
detected in \cite{Cason-Friedman-Hopkins-2014} if action updating was
performed in discrete time.
Here and in Ref.~\cite{Xu-Zhou-Wang-2013}
we find that  even discrete-time updating of actions will lead to
collective cyclic motions in the RPS game, and such a
population-level behavior is not affected by the 
particular value of $a$.

\begin{table}[t]
  \title{{\bf Table 1.}
Empirical cycling frequencies $f_{1,300}$ of $59$ populations.}
  \begin{center}
    \begin{tabular}{rrrrrr}
      \hline
      & $a=1.1$       & $2$         & $4$        & $9$        & $100$       \\
      & $f_{1,300}$  & $f_{1,300}$  & $f_{1,300}$ & $f_{1,300}$ & $f_{1,300}$  \\
      \hline
      & $0.039$     & $0.019$     & $0.033$    & $0.007$    & $0.047$      \\
      & $0.023$     & $0.023$     & $0.005$    & $-0.002$   & $0.004$      \\
      & $0.005$     & $0.054$     & $0.029$    & $0.053$    & $0.024$      \\
      & $0.029$     & $0.034$     & $0.041$    & $0.027$    & $0.051$      \\
      & $0.015$     & $-0.010$    & $0.008$    & $0.068$    & $0.027$      \\
      & $0.052$     & $0.052$     & $0.042$    & $-0.017$   & $0.031$      \\
      & $0.028$     & $0.084$     & $0.069$    & $0.032$    & $0.017$      \\
      & $0.034$     & $0.041$     & $-0.022$   & $0.049$    & $-0.017$     \\
      & $0.073$     & $-0.013$    & $0.069$    & $0.020$    & $-0.012$     \\
      & $0.023$     & $0.017$     & $0.035$    & $-0.022$   & $0.053$      \\
      & $0.018$     & $-0.005$    & $0.048$    & $0.018$    & $-0.010$     \\
      &             & $0.028$     & $0.018$    & $0.032$    & $-0.005$     \\
      \hline
      $\mu$ & $0.031$  & $0.027$  & $0.031$    & $0.022$    & $0.018$      \\
      $\sigma$ & $0.019$ & $0.029$ & $0.026$   & $0.027$    & $0.025$      \\
      $\delta$ & $0.006$ & $0.008$ & $0.008$   & $0.008$    & $0.007$      \\
      \hline
    \end{tabular}
      \begin{flushleft}
        $\mu$: the mean cycling frequency, $\sigma$: the
        standard deviation (s.d.) of cycling frequencies,
        $\delta$:  the standard error
        (SEM) of the mean cycling frequency ($\delta = \sigma/\sqrt{n_s}$,
        with sample number  $n_{s}=11$ for $a=1.1$ and
        $n_{s}=12$ for $a=2$, $4$, $9$ and $100$).
      \end{flushleft}
  \end{center}
\end{table}

\subsection*{Empirical conditional response patterns}
Under the assumption of mixed-strategy NE, the social
state transitions should obey the detailed balance condition. Therefore
the observed persistent cycling behavior cannot be understood
within  the NE framework. 
Persistent cycling can also not be explained by  the
independent decision model
which assumes the action
choice of a player at one time is influenced only by
her action at the previous time (see Supporting Information).
Using the empirically determined action shift probabilities
of  Fig.~S1 as inputs, we find that this independent
decision model
predicts the cycling frequency to be $0.0050$ (for $a=1.1$),
$-0.0005$ ($a=2.0$), $-0.0024$ ($a=4.0$), $-0.0075$ ($a=9.0$)
and $-0.0081$ ($a=100.0$), which are all very close to zero and
significantly different from the empirical values.

\begin{figure*}[t]
  \begin{center}
    \centerline{\includegraphics[width=0.55\textwidth,angle=270]{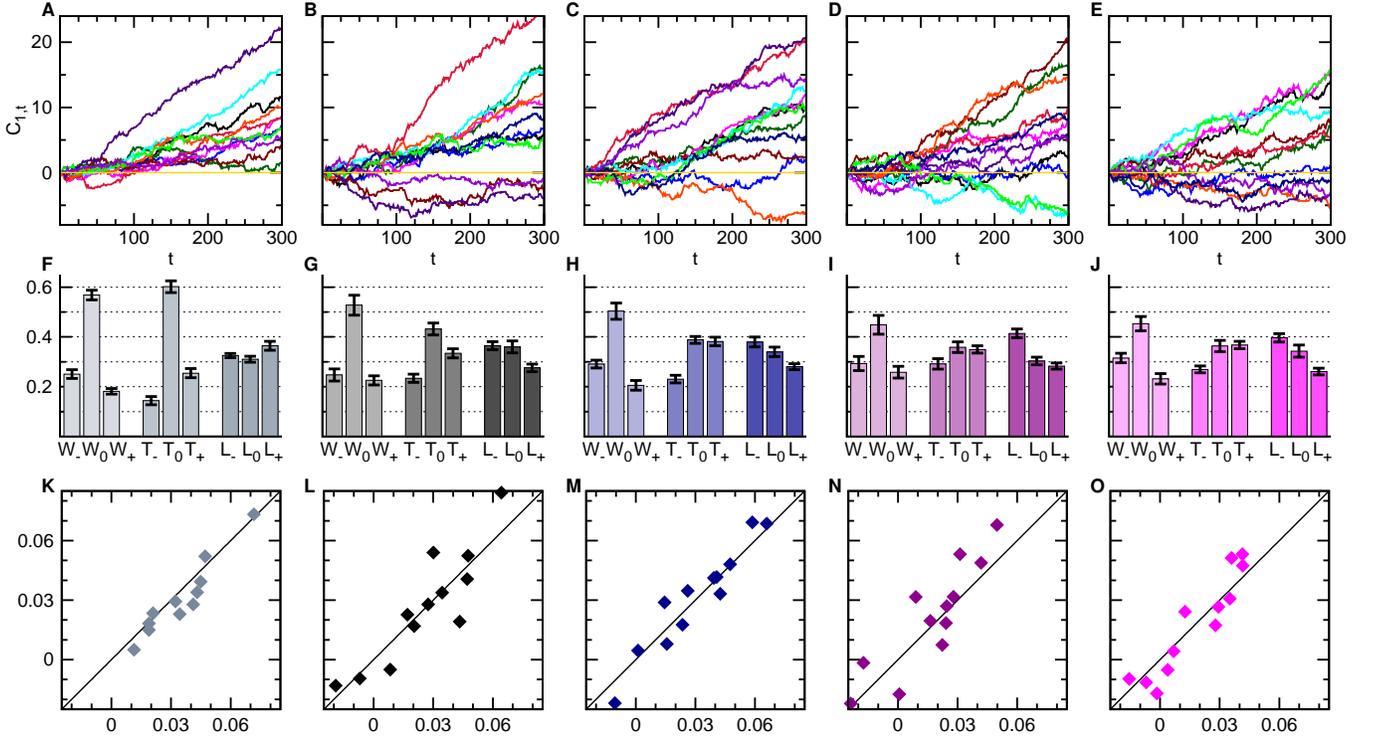}}
    \caption{
      Social cycling explained by conditional response.
      The payoff parameter is
      $a=1.1$, $2$, $4$, $9$ and $100$ from left-most column
      to right-most column.
      (A-E) Accumulated cycle numbers $C_{1, t}$
      of $59$ populations.
      (F-J) Empirically determined CR
      parameters, with the mean (vertical bin) and the
      SEM (error bar) of each CR parameter
      obtained by considering all the
      populations of the same $a$ value.
      (K-O) Comparison between the empirical cycling
      frequency (vertical axis)
      of each population and the
      theoretical frequency (horizontal axis)
      obtained by using the empirical CR parameters of this
      population as inputs.}
    \end{center}
\end{figure*}

The action choices of different players must be mutually
influenced. Our empirical data shown in Fig.~2 (F--J) confirm the existence
of such mutual influences.
Let us denote by $O$ the performance (output) of a player at a given game round, with
$O \in \{W\; ({\rm win}), T \; ({\rm tie}), L \; ({\rm lose})\}$.
Conditional on the output $O$, the probability that this player will decide to
shift action clockwise or counter-clockwise or keep the same action in the
next play is denoted as $O_{-}$, $O_{+}$ and $O_{0}$
($\equiv 1-O_{-}-O_{+}$), respectively.
Most interestingly, we see from Fig.~2 (F--J) that if a player wins over
her opponent in one play, her probability ($W_0$) of repeating the same
action in the next play is considerably higher than her probabilities
($W_{-}$ and $W_{+}$) of shifting actions. Furthermore, for payoff parameter $a\geq 2$,
if a player loses to her opponent in one play, she is more likely to shift action
clockwise (probability $L_{-}$) than either to keep the old action ($L_{0}$) or to 
shift action counter-clockwise ($L_{+}$).

\subsection*{The conditional response model}
Inspired by these empirical observations, we develop a simplest
nontrival model by assuming the following conditional response strategy:
at each game round, every player review her previous performance 
$O \in \{W, T, L\}$ and makes an action choice according to the
corresponding three conditional probabilities ($O_{-}$, $O_{0}$, $O_{+}$).
This model is characterized by a set
$\Gamma \equiv \{ W_{-}, W_{+}; T_{-},
T_{+}; L_{-}, L_{+}\}$ of six CR  parameters.
Notice this CR model differs qualitatively from
 the discrete-time logit dynamics model \cite{Blume-1993,Hommes-Ochea-2012} used in
Ref.~\cite{Xu-Zhou-Wang-2013}, which assumes each
player has global information about the population's social state.

We can solve this win-lose-tie CR model analytically and numerically. 
Let us denote by $n_{r r}$, $n_{p p}$, $n_{s s}$, 
$n_{r p}$, $n_{p s}$ and $n_{s r}$, respectively,
as the number of pairs in which the competition being $R$--$R$,
$P$--$P$,  $S$--$S$, $R$--$P$, $P$--$S$, and $S$--$R$, in one
game round $t$. 
Given the social state ${\bf s}= (n_R, n_P, n_S)$ at time $t$, the 
conditional joint probability distribution of these
six integers is expressed as
\begin{eqnarray}
& & \hspace*{-1.0cm}P_{{\bf s}}\bigl[ n_{r r}, n_{p p}, n_{s s}, n_{r p}, n_{p s} , n_{s r}
\bigr]  = \nonumber \\
& &  \hspace*{-0.75cm}
\frac{
n_{R}! n_{P}! n_{S}! 
\delta_{2 n_{r r} + n_{s r} + n_{r p}}^{n_{R}} 
\delta_{2 n_{p p} + n_{r p} + n_{p s}}^{n_{P}} 
\delta_{2 n_{s s} + n_{p s} + n_{s r}}^{n_{S}}}
{ (N-1)!!  2^{n_{r r}} n_{r r}!  2^{n_{p p}} n_{p p}! 
2^{n_{s s}}  n_{s s}! n_{r p}! n_{p s}!  n_{s r}!} \; ,
\label{eq:PairProbcr}
\end{eqnarray}
where $(N-1)!! \equiv
1\times 3 \times \ldots \times (N-3) \times (N-1)$ and
$\delta_{m}^{n}$ is the Kronecker symbol
($\delta_{m}^{n}=1$ if $m=n$ and $=0$ if otherwise).
With the help of this expression,
we can then obtain an explicit formula for the 
social state transition probability 
$M_{c r}[{\bf s}^\prime | {\bf s}]$ from ${\bf s}$ to any another
social state ${\bf s}^\prime$. We
then compute numerically the 
steady-state social state distribution $P_{c r}^*({\bf s})$
of this Markov matrix 
\cite{Kemeny-Snell-1983}
and other average quantities of interest.
For example, the mean steady-state cycling frequency $f_{c r}$ of this
model is computed by
\begin{equation}
f_{cr} = \sum\limits_{{\bf s}} P_{cr}^{*}({\bf s})
\sum\limits_{{\bf s}^\prime} M_{cr}[{\bf s}^\prime | {\bf s}]
\theta_{{\bf s}\rightarrow {\bf s}^\prime} \; ,
\end{equation}
where $\theta_{{\bf s}\rightarrow {\bf s}^\prime}$ is the
rotation angle associated with the social state transition
${\bf s}\rightarrow {\bf s}^\prime$, see Eq.~[\ref{eq:thetaval}].

Using the empirically determined response parameters as
inputs, the CR model predicts the mean cycling frequencies
for the five sets of populations to be 
$f_{c r}= 0.035$ ($a=1.1$), $0.026$ ($a=2$),
$0.030$ ($a=4$), $0.018$ ($a=9$) and $0.017$ ($a=100$),
agreeing well with the empirical measurements.
Such good agreements between model and experiment
are achieved also for the $59$ individual populations
(Fig.~2 K--O).

Because of the rotational symmetry of the conditional response
parameters, the CR model predicts that each player's action marginal
distribution is uniform (see Supporting Information), 
identical to the NE mixed strategy. On the other hand, according to this
model, the  expected payoff $g_{c r}$  per game round of each player is
\begin{equation}
g_{cr}  = g_{0} + (a-2) \times (1/6  - \tau_{c r}/2 ) \; ,
\end{equation}
where $g_0\equiv (1+a)/3$ is the expected payoff of the NE mixed
strategy, and $\tau_{c r}$ is the average fraction of ties among the
$N/2$ pairs at each game round, with the expression
\begin{equation}
\tau_{c r} = 
\sum\limits_{{\bf s}}
P_{cr}^{*}({\bf s}) \frac{
 \sum\limits_{n_{r r},
 \ldots , n_{r s}}  (n_{r r} + n_{p p} + n_{s s})
 P_{{\bf s}}(n_{r r}, \ldots, n_{s r})}{N/2} \; .
\end{equation}
The value of $g_{c r}$ depends on the CR parameters.
By uniformly sampling $2.4\times 10^9$ instances of $\Gamma$
from the three-dimensional probability
simplex,
we find that for $a>2$, $g_{cr}$ has high chance of being
lower than $g_0$  (Fig.~3), with the mean value of
$(g_{cr}-g_0)$ being $-0.0085 (a-2)$.  
(Qualitatively the same conclusion is obtained for larger $N$ values, e.g.,
see Fig.~S2 for $N=12$.) 
This is consistent with the mixed-strategy
NE being evolutionarily stable \cite{Taylor-Jonker-1978}.
On the other hand, the five $g_{cr}$ values
determined by the empirical CR parameters and the
corresponding five mean payoffs of the
empirical data sets all weakly exceed $g_0$,
indicating that individual players are adjusting their
responses to achieve higher accumulated payoffs (see Supporting
Information).
The positive gap between $g_{c r}$ and $g_0$ may further
enlarge if the individual players were given more learning time to optimize their
response parameters (e.g., through increasing the repeats of the game).

As shown in Fig.~3 and Fig.~S2, the CR parameters have to be
highly optimized to achieve a large value of $g_{c r}$.
For population size $N=6$
we give three examples of the sampled best CR strategies for $a>2$:
$\Gamma_1=\{0.002, 0.000; 0.067, 0.110; 0.003, 0.003\}$,
with cycling frequency $f_{cr}=0.003$ and
$g_{cr}=g_0+0.035 (a-2)$;
$\Gamma_2=\{0.995, 0.001; 0.800, 0.058;
0.988, 0.012\}$, with $f_{cr}=-0.190$ and
$g_{cr}=g_0 + 0.034 (a-2)$;
$\Gamma_3=\{0.001, 0.004; 0.063, 0.791;
0.989, 0.001\}$, with $f_{cr}=0.189$ and
$g_{cr}=g_0+0.033 (a-2)$.
For large $a$ these CR strategies outperform the
NE mixed strategy in payoff by about $10\%$.
Set $\Gamma_1$ indicates that
population-level cycling is not a necessary
condition for achieving high payoff values.
On the other hand, set $\Gamma_3$
implies $W_0\approx 1, L_0 \approx 0$,
therefore this CR strategy can be regarded as an extension
of the win-stay lose-shift
(also called Pavlov) strategy,
which has been shown by computer simulations to facilitate
cooperation in the prisoner's dilemma game
\cite{Kraines-Kraines-1993,Nowak-Sigmund-1993,Wedekind-Milinski-1996,Posch-1999}.

\begin{figure}[t]
  \begin{center}
    \centerline{\includegraphics[width=0.325\textwidth,angle=270]{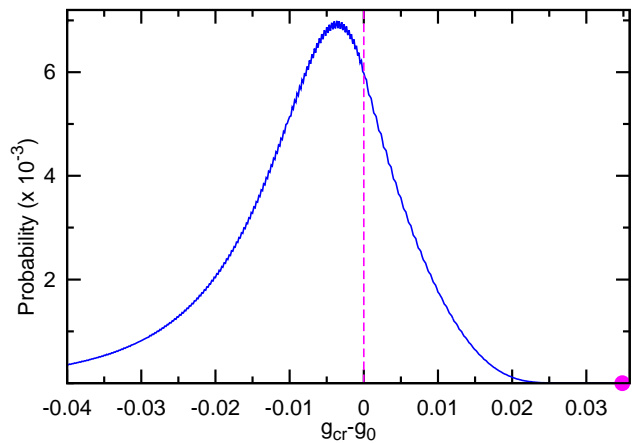}}
    \caption{
      Probability distribution of payoff difference
      $g_{cr}-g_{0}$ at population size $N=6$.
      We assume $a>2$ and set the unit of the horizontal axis to
      be $(a-2)$. 
      The solid line is obtained by sampling
      $2.4\times 10^9$ CR strategies uniformly at random;
      the filled circle denotes the maximal value of
      $g_{cr}$ among these samples.}
  \end{center}
\end{figure}


\section*{Discussion}

In game-theory literature it is common to equate individual
players' action marginal distributions
with their actual strategies\cite{Sandholm-2010,Osborne-Rubinstein-1994}.
In reality, however, decision-making and learning are very
complicated neural
processes
\cite{Glimcher-etal-2009,Borgers-Sarin-1997,Posch-1997,Galla-2009}.
The action marginal distributions are
only a consequence of such complex dynamical processes,
their coarse-grained nature makes them unsuitable to describe
dynamical properties.
Our work on the finite-population RPS game
clearly demonstrates this point.
This game exhibits collective cyclic motions which
cannot be understood by the NE concept but are successfully
explained by the empirical data-inspired CR
mechanism. As far as the action marginal distributions of
individual players are concerned, the CR strategy is
indistinguishable from the NE mixed strategy, yet
it is capable of bringing  higher payoffs to
the players if its parameters are optimized.
This simple conditional response strategy, with the win-stay lose-shift
strategy being a special case, appears to be psychologically
plausible for human subjects with bounded rationality \cite{Camerer-1999,Camerer-2003}.
For more complicated game payoff matrices, we can
generalize the conditional response model accordingly
by introducing a larger set of CR parameters.
It should be very interesting to re-analyze many existing
laboratory experimental data
\cite{Berninghaus-Ehrhart-Keser-1999,Traulsen-etal-2010,Hoffman-etal-2012,GraciaLazaro-etal-2012,Frey-Goldstone-2013,Cason-Friedman-Hopkins-2014,Chmura-Goerg-Selten-2014}
using this extended model.

The CR model as a simple model of decision-making
under uncertainty deserves to be fully explored.
We find the
cycling frequency is not sensitive
to population size $N$ at given
CR parameters (see Fig.~S3); and the cycling frequency
is nonzero even for symmetric CR parameters
(i.e., $W_{+}/W_{-} = T_{+}/T_{-} =
L_{+} / L_{-}=1$), as long as $W_{0} \neq L_{0}$ (see Fig.~S4).
The optimization issue of CR parameters is
left out in this work. We will investigate whether an
optimal CR strategy is achievable through
simple stochastic learning
rules \cite{Borgers-Sarin-1997,Posch-1997}.
The effects of memory length \cite{Press-Dyson-2012} and
population size to the optimal CR strategies
also need to be thoroughly studied.
On the more biological side, whether CR is a basic
decision-making mechanism of the human brain or just 
a consequence of more fundamental neural mechanisms
is a challenging question for future studies.

\section*{Materials and Methods}

{\bf Experiment.} The experiment was performed at Zhejiang
University in the period of December 2010 to March 2014.
A total number of $360$ students of Zhejiang University
volunteered to serve as the human subjects of this
experiment. Informed consent was obtained from all the  participants. 
These human subjects were distributed to $60$ populations of equal size
$N=6$. The subjects of each population played within themselves the
RPS game for $300$ rounds under the random pairwise-matching protocol
(see Supporting Information for additional details), with the
payoff parameter $a$ fixed to one of five different values.
After the RPS game each human subject was rewarded by cash (RMB) privately. 
Suppose the accumulated payoff of a human subject is $x$ in the game,
then the reward $y$ in RMB is
$y =  r \times x   + 5$,
where the exchange rate $r$ depends on $a$. According to the
Nash equilibrium theory, the expected payoff of each player in one game
round is $(1+a)/3$. Therefore we set $r=0.45/(1+a)$, so that the expected reward
in RMB to each human subject will be the same ($=50$ RMB) for all the
$60$ populations. The numerical value of $r$ and the reward formula were
both informed to the human subjects before the RPS game.

{\bf Rotation angle computation.}
Consider a transition from one social state ${\bf s}= (n_R, n_P,
n_S)$ at game round $t$ to another social state
$\tilde{{\bf s}} = (\tilde{n}_R, \tilde{n}_P, \tilde{n}_S)$ at
game round $(t+1)$, 
if at least one of the two social states coincides
with the centroid ${\bf c}_0$ of the social state plane,
or the three points ${\bf s}$, $\tilde{\bf s}$ and
${\bf c}_0$ lie on a straight line, then the transition
${\bf s}\rightarrow \tilde{\bf s}$ is not regarded as a rotation
around ${\bf c}_0$, and the rotation angle $\theta=0$.
In all the other cases, the transition ${\bf s}\rightarrow
\tilde{\bf s}$ is regarded as a rotation around ${\bf c}_0$,
and the rotation angle is computed through
\begin{eqnarray}
  & & \hspace*{-0.7cm}\theta =
  {\rm sgn}_{{\bf s} \rightarrow \tilde{\bf s}}
  \times  \nonumber \\
  & & \hspace*{-0.5cm}{\rm acos}\Bigl(
  \frac{3 (n_R \tilde{n}_R + n_P \tilde{n}_P + n_S
    \tilde{n}_S)-N^2}{\sqrt{
      [3 (n_R^2 + n_P^2 +n_S^2)-N^2]
      [3 (\tilde{n}_R^2+ \tilde{n}_P^2+
        \tilde{n}_S^2)-N^2]}}\Bigr) \; ,
  \nonumber \\
  & & \label{eq:thetaval}
\end{eqnarray}
where ${\rm acos}(x)\in [0, \pi)$ is the inverse cosine function, and
  ${\rm sgn}_{{\bf s} \rightarrow \tilde{\bf s}} = 1$
  if $[3(n_R \tilde{n}_P- n_P \tilde{n}_R )+ N (n_P - n_R
    +\tilde{n}_R - \tilde{n}_P)]>0$ (counter-clockwise rotation
  around ${\bf c}_0$)
  and ${\rm sgn}_{{\bf s} \rightarrow \tilde{\bf s}}
  = -1$ if otherwise (clockwise rotation around ${\bf c}_0$).
  
  {\bf Statistical Analysis.}
  Statistical analyses, including Wilcoxon
  signed-rank test and Spearman's rank correlation test,
  were performed by using stata 12.0  (Stata, College Station, TX).

  \section*{Acknowledgments}
  
  ZW and BX were supported by the Fundamental Research Funds for the Central 
  Universities
  (SSEYI2014Z), the State Key Laboratory for Theoretical
  Physics (Y3KF261CJ1),
  and the Philosophy and Social
  Sciences Planning Project of Zhejiang Province (13NDJC095YB);
  HJZ was supported by the National Basic Research Program
  of China (2013CB932804),the Knowledge Innovation Program of
  Chinese Academy of Sciences (KJCX2-EW-J02),
  and the National Science Foundation of China
  (11121403, 11225526).
  We thank Erik Aurell and Angelo Valleriani for helpful
  comments on the manuscript.


\clearpage

\widetext 

\begin{center}
{\large Supplementary Table 1}
\end{center}
\vskip 0.5cm

\begin{center}
{\bf Table S1.} 
Statistics on individual players' action marginal probabilities.
\vskip 0.5cm
 \begin{tabular}{rrcrrrr}
    \hline
     $a$  &  \quad $m$ & \ action \
    &\ \  $\mu$   & \ \ \quad $\sigma$ &  \ \ Min & \ \  Max  \\
    \hline
    $1.1$ & $66$ & $R$ & $0.37$  & $0.08$ & $0.19$ &
    $0.68$ \\
    & & $P$ & $0.34$ & $0.07$ & $0.18$ & $0.52$ \\
    & & $S$ & $0.30$ & $0.06$ & $0.09$ & $0.41$ \\
    \hline
    $2$ & $72$ & $R$ & $0.36$ & $0.07$ & $0.14$ &
    $0.60$ \\
    & & $P$ & $0.32$ & $0.07$ & $0.15$ & $0.58$ \\
    & & $S$ & $0.32$ & $0.06$ & $0.13$ & $0.46$ \\
    \hline
    $4$ & $72$ & $R$ & $0.35$ & $0.08$ & $0.11$
    & $0.60$ \\
    &  & $P$ & $0.33$ & $0.07$ & $0.14$ & $0.54$\\
    & & $S$ & $0.32$ & $0.07$ & $0.11$ & $0.50$ \\
    \hline
    $9$ & $72$ & $R$ & $0.35$ & $0.08$ & $0.21$ &
    $0.63$ \\
    & & $P$ & $0.33$ & $0.07$ & $0.13$ & $0.55$ \\
    & & $S$ & $0.32$ & $0.06$ & $0.16$ & $0.53$ \\
    \hline
    $100$ & $72$ & $R$ & $0.35$ & $0.07$ & $0.22$ &
    $0.60$ \\
    & & $P$ & $0.33$ & $0.05$ & $0.16$ & $0.51$ \\
    & & $S$ & $0.32$ & $0.06$ & $0.14$ & $0.47$ \\
    \hline
     & $354$ & $R$ & $0.36$ & $0.08$ & $0.11$ & $0.68$ \\
    & & $P$ & $0.33$ & $0.07$ & $0.13$ & $0.58$ \\
    & & $S$ & $0.32$ & $0.06$ & $0.09$ & $0.53$ \\
    \hline
 \end{tabular}
 \vskip 0.5cm
 \begin{flushleft}
   $m$ is the total number of players; $\mu$, $\sigma$,
   Max and Min are, respectively, the mean,
   the standard deviation (s.d.), the maximum and
   minimum of the action marginal probability
   in question among all the $m$ players. The
   last three rows are statistics performed on all the
   $354$ players.
 \end{flushleft}
\end{center}

\clearpage

\begin{center}
  {\large Supplementary Table 2}
\end{center}
\vskip 0.5cm

\begin{center}
  {\bf Table S2.} Empirical cycling frequencies $f_{1,150}$ and
  $f_{151,300}$ for $59$ populations.
\vskip 0.5cm
\begin{tabular}{rrrrrrrrrrr}
      \hline
      & & $1.1$  
      & & $2$    	
      & & $4$    
      & & $9$    
      & & $100$  
      \\
       & $f_{1,150}$ & $f_{151,300}$
       &  $f_{1,150}$ & $f_{151,300}$
       & $f_{1,150}$ & $f_{151,300}$
       & $f_{1,150}$ & $f_{151,300}$
       & $f_{1,150}$ & $f_{151,300}$
      \\
      \hline
      & $0.032$  & $0.047$
      & $0.020$  & $0.017$ 	
      & $0.016$  & $0.050$
      & $-0.007$ & $0.022$
      & $0.040$  & $0.052$
      \\
      & $0.008$  & $0.039$ 	
      & $0.028$  & $0.017$
      & $-0.002$ & $0.014$
      & $-0.005$ & $0.001$
      & $-0.003$ & $0.009$
     \\
      & $0.025$  & $-0.014$
      & $0.021$  & $0.087$
      & $0.023$  & $0.035$
      & $0.044$  & $0.062$
      & $0.009$  & $0.038$
      \\
      & $0.015$  & $0.045$
      & $0.023$  & $0.043$
      & $0.024$  & $0.059$
      & $0.034$  & $0.020$
      & $0.045$  & $0.060$
      \\
      & $0.011$  & $0.019$
      & $-0.027$ & $0.006$
      & $0.019$  & $-0.004$
      & $0.045$  & $0.088$
      & $0.017$  & $0.038$
      \\
      & $0.036$  & $0.068$
      & $0.024$  & $0.081$
      & $0.018$  & $0.068$
      & $-0.022$ & $-0.014$
      & $0.055$  & $0.006$
      \\
      & $0.010$ & $0.045$
      & $0.083$ & $0.086$
      & $0.079$ & $0.059$
      & $0.032$ & $0.030$
      & $0.008$ & $0.026$
      \\
      & $0.036$  & $0.033$
      & $0.034$  & $0.046$
      & $-0.013$ & $-0.031$
      & $0.047$  & $0.050$
      & $-0.019$ & $-0.015$
      \\
      & $0.076$  & $0.070$
      & $-0.032$ & $0.004$
      & $0.077$  & $0.061$
      & $0.010$  & $0.027$
      & $-0.034$ & $0.010$
      \\
      & $0.029$ & $0.016$
      & $0.034$ & $-0.002$
      & $0.018$ & $0.051$
      & $-0.003$ & $-0.041$
      & $0.055$ & $0.052$
      \\
      & $0.009$  & $0.025$
      & $-0.004$ & $-0.006$
      & $0.061$  & $0.038$
      & $0.005$ & $0.029$
      & $-0.012$ & $-0.007$
      \\
      &         &
      & $0.022$ & $0.031$
      & $0.017$ & $0.019$
      & $0.011$ & $0.055$
      & $-0.017$ & $0.007$
      \\
      \hline
      $\mu$	
      & $0.026$ & $0.036$
      & $0.019$ & $0.034$
      & $0.028$ & $0.035$
      & $0.016$ & $0.027$
      & $0.012$ & $0.023$
      \\
      $\sigma$
      & $0.020$ & $0.024$
      & $0.030$ & $0.035$
      & $0.029$ & $0.030$
      & $0.024$ & $0.035$
      & $0.031$ & $0.025$
      \\
      $\delta$ 
      & $0.006$ & $0.007$
      & $0.009$ & $0.010$
      & $0.008$ & $0.009$
      & $0.007$ & $0.010$
      & $0.009$ & $0.007$
      \\
      \hline
\end{tabular}
\vskip 0.5cm
\begin{flushleft}
  The first row shows the value of the payoff parameter $a$.
  For each experimental session (population),
  $f_{1,150}$ and $f_{151,300}$ are respectively
  the cycling frequency in the first and the
  second $150$ time steps.
  $\mu$ is the mean cycling frequency, $\sigma$ is the
  standard deviation (s.d.) of the cycling frequency,
  $\delta = \sigma / \sqrt{n_s}$ is the standard error
  (SEM) of the mean cycling frequency. The number of
  populations is $n_{s}=11$ for $a=1.1$ and
  $n_{s}=12$ for $a=2$, $4$, $9$ and $100$.
\end{flushleft}
\end{center}

\clearpage

\begin{center}
  {\large Supplementary Figure 1}
\end{center}
\vskip 0.2cm

\begin{center}
  \psfrag{Rm}{\hspace{-0.05cm}{\scriptsize $R_{-}$}}
  \psfrag{R0}{{\scriptsize $R_{0}$}}
  \psfrag{Rp}{\hspace{0.05cm}{\scriptsize $R_{+}$}}
  \psfrag{Pm}{\hspace{-0.05cm}{\scriptsize $P_{-}$}}
  \psfrag{P0}{{\scriptsize $P_{0}$}}
  \psfrag{Pp}{\hspace{0.05cm}{\scriptsize $P_{+}$}}
  \psfrag{Sm}{\hspace{-0.05cm}{\scriptsize $S_{-}$}}
  \psfrag{S0}{{\scriptsize $S_{0}$}}
  \psfrag{Sp}{\hspace{0.05cm}{\scriptsize $S_{+}$}}
  \psfrag{0.2}{\hspace{-0.1cm}{\footnotesize $0.2$}}
  \psfrag{0.4}{\hspace{-0.1cm}{\footnotesize $0.4$}}
  \psfrag{0.6}{\hspace{-0.1cm}{\footnotesize $0.6$}}
  \includegraphics[scale=0.5,angle=270]{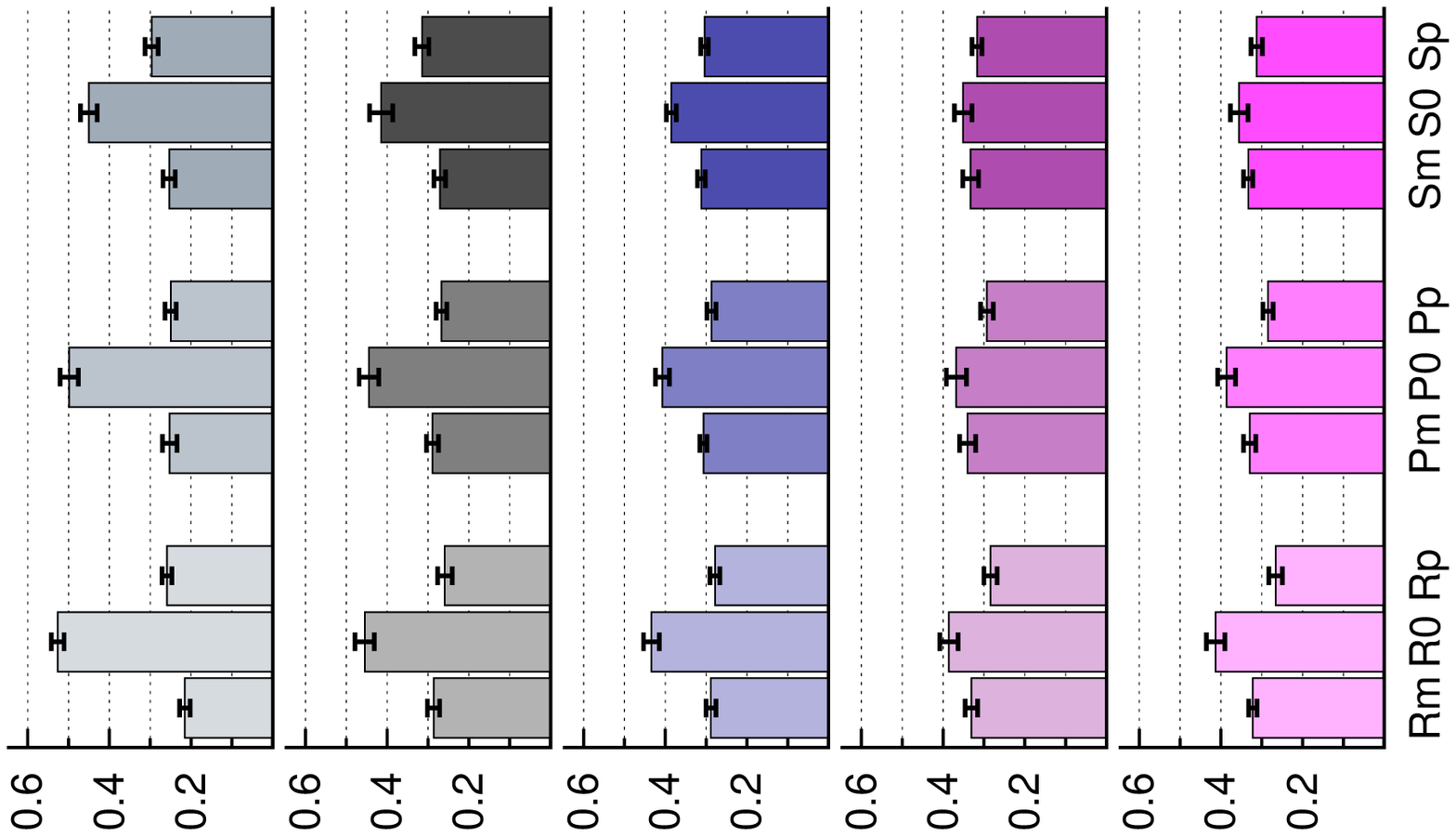}
\end{center}
\vskip 0.2cm
\begin{flushleft}
{\bf Figure S1.} Action shift probability conditional on a player's
    current action.
    If a player adopts the action $R$ at one game round,
    this player's conditional probability of repeating the
    same action at the next game round is denoted as $R_{0}$,
    while the conditional probability of performing a
    counter-clockwise or clockwise action shift is denoted,
    respectively, as $R_{+}$ and $R_{-}$. The conditional
    probabilities $P_{0}, P_{+}, P_{-}$ and $S_{0},
    S_{+}, S_{-}$ are defined similarly. The mean value
    (vertical bin) and the SEM (standard error of the mean, 
    error bar) of each conditional probability is obtained by averaging over 
    the different populations of the
    same value of $a=1.1$, $2$, $4$, $9$, and $100$
    (from top row to bottom row).
\end{flushleft}

\clearpage

\begin{center}
{\large Supplementary Figure 2}
\end{center}
\vskip 0.2cm

 \begin{center}
   \includegraphics[scale=1.0,angle=270]{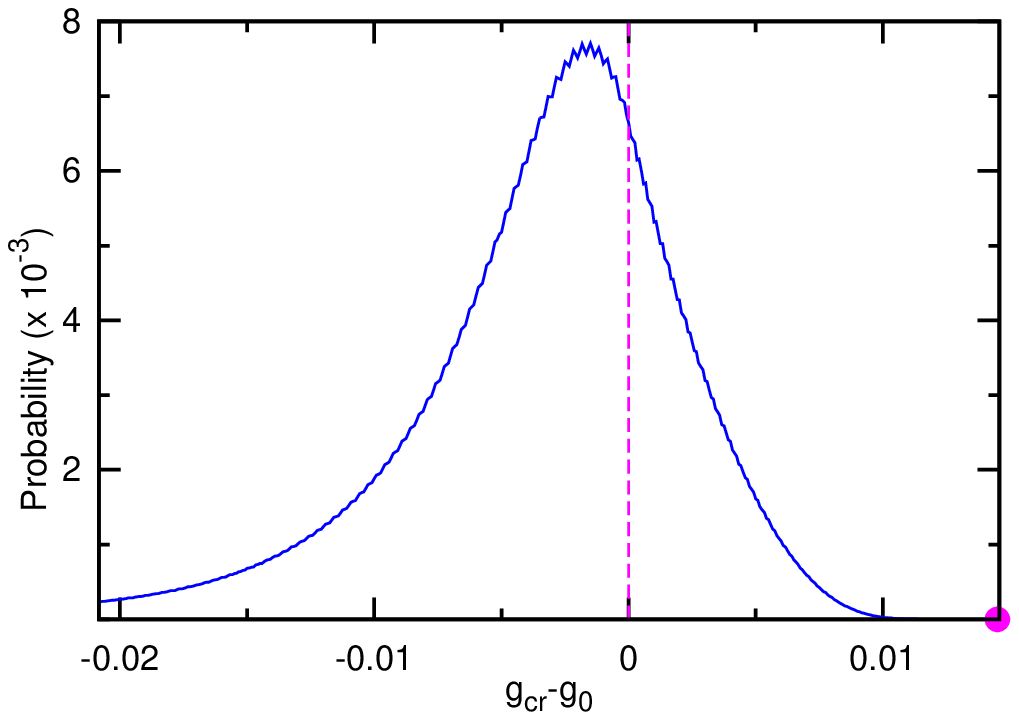}
\end{center}
 \vskip 0.2cm
\begin{flushleft}
  {\bf Figure S2.}
  Probability distribution of payoff difference
  $g_{cr}-g_{0}$ at population size $N=12$.
  As in Fig.~3, we assume $a>2$ and set the unit of the horizontal axis to
  be $(a-2)$. 
  The solid line is obtained by sampling
  $2.4\times 10^9$ CR strategies uniformly at random;
  the filled circle denotes the maximal value of
  $g_{cr}$ among these samples.
\end{flushleft}

\clearpage

\begin{center}
{\large Supplementary Figure 3}
\end{center}
\vskip 0.2cm

\begin{center}
  \includegraphics[scale=0.75,angle=270]{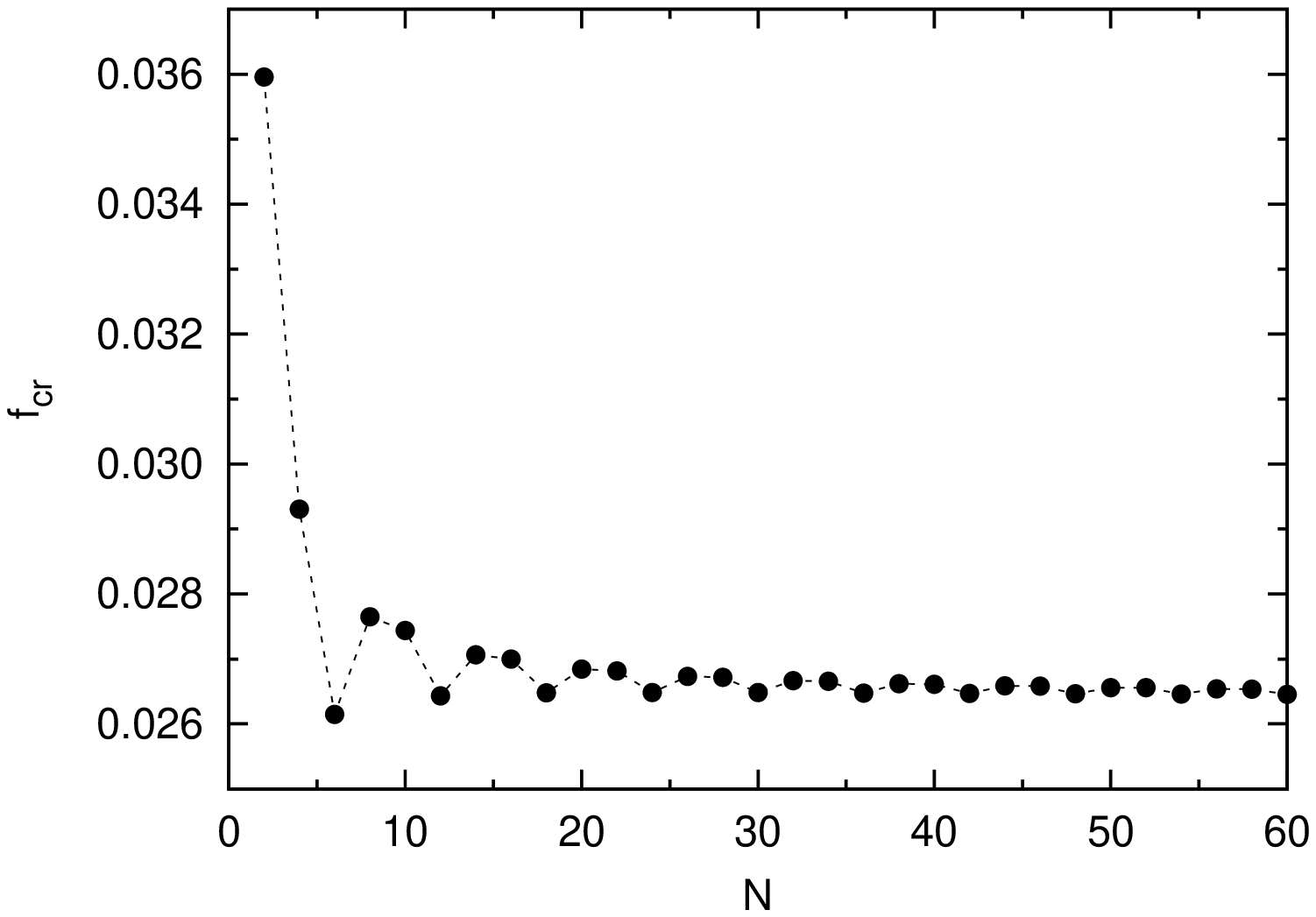}
\end{center}
\vskip 0.2cm
\begin{flushleft}
  {\bf Figure S3.}
  The cycling frequency $f_{c r}$ of the conditional response model as a
  function of population size $N$.
  For the purpose of illustration, the CR parameters shown in
  Fig.~2G are used in the numerical computations.
\end{flushleft}

\clearpage

\begin{center}
{\large Supplementary Figure 4}
\end{center}
\vskip 0.2cm

 \begin{center}
   \includegraphics[scale=0.65,angle=270]{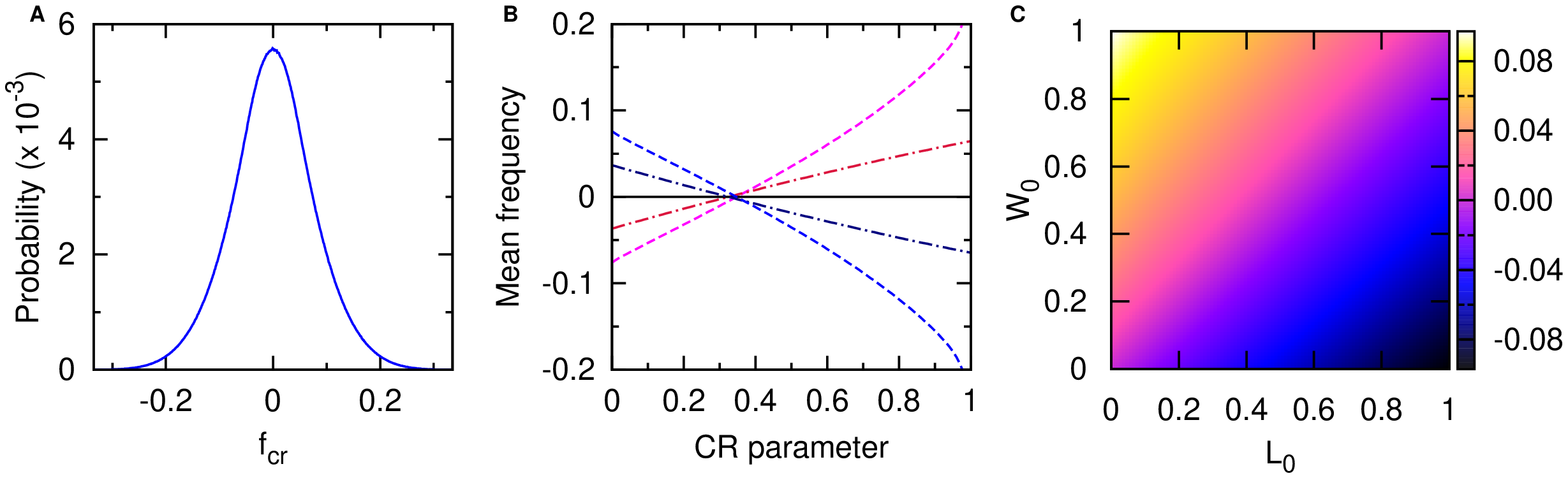}
\end{center}
\vskip 0.2cm
\begin{flushleft}
  {\bf Figure S4.}
  Theoretical predictions of the conditional response model with
  population size $N=6$.
  (A) Probability distribution of the cycling frequency $f_{c r}$
  obtained by sampling $2.4\times 10^9$ CR strategies uniformly at
  random. (B) The mean value of $f_{c r}$ as a function of one
  fixed CR parameter while the remaining CR parameters are sampled
  uniformly at random. The fixed CR parametr is $T_{+}$ (red dashed
  line), $W_{0}$ or $L_{+}$ (brown dot-dashed line), $W_{+}$ or
  $T_{0}$ or $L_{-}$ (black solid line), $W_{-}$ or $L_{0}$ (purple
  dot-dashed line), and $T_{-}$ (blue dashed line).
  (C) Cycling frequency $f_{c r}$ as a function of CR parameters
  $W_{0}$ and $L_{0}$ for the symmetric 
  CR model ($W_{+}/W_{-}=T_{+}/T_{-}=L_{+}/L_{-}=1$) with $T_0=0.333$.
\end{flushleft}

\clearpage

\endwidetext

\begin{center}
{\Large Supporting Information}
\end{center}


\section{Experimental setup}

We carried out five sets of experimental sessions
at different days during
the period of December 2010 to March 2014, with
each set consisting of $12$ individual experimental
sessions. The payoff parameter value was fixed to
$a=1.1$, $2$, $4$, $9$ and $100$,
respectively, in these five sets of experimental sessions.
Each experimental session involved $N=6$ human subjects (players) and it was carried out at Zhejiang University
within a single day.

We recruited a total number of $72 \times 5 = 360$
undergraduate and graduate students from
different disciplines of Zhejiang University.
These students served as the players of our
experimental sessions, each of which participating in
only one experimental session.
Female students were more enthusiastic than
male students in registering
as candidate human subjects of our experiments.
As we sampled students uniformly at random from the candidate
list, therefore more female students were recruited than
male students (among the $360$ students, the female versus male
ratio is $217 : 143$).
For each set of experimental sessions,
the recruited $72$ players were distributed
into $12$ groups
(populations) of size $N=6$ uniformly at random by a
computer program.

The players then sited separately in a classroom, each of
which facing a computer screen. They were not
allowed to communicate with each other during the whole
experimental session.
Written instructions were handed out to each player
and the rules of the experiment were also
orally explained by an experimental instructor.
The rules of the experimental session are as follows:
\begin{enumerate}
\item[(i)]
 Each player plays the Rock-Paper-Scissors (RPS) game
repeatedly with the same other five players for a total
number of $300$ rounds.

\item[(ii)] Each player earns virtual points during
the experimental session according to the payoff matrix shown
in the written instruction. These virtual points are
then exchanged into RMB as a reward to the player,
plus an additional $5$ RMB as show-up fee. (The
exchange rate between virtual point and RMB is
the same for all the $72$ players of these
$12$ experimental sessions. Its actual value is
informed to the players.)

\item[(iii)] In each game round, the six players of
each group are randomly matched by a computer program
to form three pairs, and each player plays the RPS game only with the assigned pair opponent.

\item[(iv)] Each player has at most $40$ seconds
in one game round to make a choice among the three candidate actions ``Rock",
 ``Paper" and ``Scissors".
If this time runs out, the player has to make a
choice immediately (the experimental instructor will
 loudly urge these players to do so).
 After a choice has been made it can not be changed.
\end{enumerate}
Before the start of the actual experimental session, the
player were asked to answer four questions
to ensure that they understand completely the
rules of the experimental session. These four questions
are:
(1)
{\em If you choose ``Rock" and your opponent chooses
``Scissors", how many virtual points will you earn?}
(2)
{\em  If you choose ``Rock" and your opponent
chooses also ``Rock", how many virtual points will you earn?}
(3)
{\em
If you choose ``Scissors" and your opponent chooses
``Rock",
how many virtual points will you earn?}
(4)
{\em
Do you know that
at each game round you will play with a randomly chosen
opponent from your group (yes/no)?}

During the experimental session, the computer screen of each
player will show an information window and a decision
window.
The window on the left of the computer screen
is the information window. The upper panel of this
information window shows the current game
round, the time limit ($40$ seconds)
of making a choice, and the
time left to make a choice. The color of this upper panel
turns to green at the start of each game round.
The color will change to yellow if the player
does not make a choice within $20$ seconds. The color will
change to red if the decision time runs out (and then
the experimental instructor will loudly urge the players to
make a choice immediately).
The color will change to blue if a choice has been made
by the player.

After all the players of the group have made their decisions,
the lower panel of the information window will show the
player's own choice, the opponent's choice, and the
player's own payoff in this game round. The player's own
accumulated payoff is also shown.
The players are asked to record their choices of each round
on the record sheet (Rock as $R$, Paper as $P$, and Scissors
as $S$).

The window on the right of the computer screen is the
decision window. It is activated only after all the players of
the group have made their choices.
The upper panel of this decision window lists the current game round, while the lower panel
lists the three candidate actions
``Rock", ``Scissors", ``Paper" horizontally from
left to right. The player can make a choice by clicking
on the corresponding action names.
After a choice has been made by the player,
the decision window becomes inactive until
the next game round starts.

The reward in RMB for each player is determined by the
following formula.
Suppose a player $i$ earns $x_i$ virtual points in the 
whole experimental session, the total reward $y_i$ in RMB for  this player is then given by  
$y_i = x_i \times r  + 5$,
where $r$ is the exchange rate between virtual point and RMB.
In this work we set $r=0.45/(1+a)$.
Then the expected total earning in RMB
for a player will be the same ($=50$ RMB) in the five sets of
experimental sessions under the assumption of mixed-strategy
Nash equilibrium, which predicts the expected payoff of each player in one game round to 
be $(1+a)/3$.
The actual numerical value of $r$ and the above-mentioned reward formula were listed in the
written instruction and also orally mentioned by the
experimental instructor at the
instruction phase of the experiment.


\section{The mixed-strategy Nash equilibrium}

The RPS game has a mixed-strategy Nash equilibrium (NE), 
in which every player of the population
adopts the three actions
($R$, $P$, and $S$) with the same probability
$1/3$ in each round of the game.
Here we give a proof of this statement.
We also demonstrate that, the empirically observed
action marginal probabilities of
individual players are consistent with the NE mixed
strategy.

Consider a population of $N$ individuals playing repeatedly
the RPS game under the random pairwise-matching protocol.
Let us define $\rho_i^R$ (respectively, $\rho_i^P$ and
$\rho_i^S$) as the probability that a player
$i$ of the population ($i\in \{1, 2, \ldots, N\}$)
will choose action $R$ (respectively, $P$ and $S$) in
one game round. If a player $j$ chooses action  $R$,
what is her expected payoff in one play?
Since this player has equal chance $1/(N-1)$ of pairing
with any another player $i$, the expected payoff is
simply $g_j^R
\equiv \sum_{i\neq j} (\rho_i^R + a \rho_i^S)/(N-1)$.
By the same argument we see that if player $j$ chooses
action $P$ and $S$ the expected payoff $g_j^P$ and
$g_j^S$ in one play are, respectively,
$g_j^P \equiv \sum_{i\neq j} (\rho_i^P + a \rho_i^R)/(N-1)$
and
$g_j^S \equiv \sum_{i\neq j} (\rho_i^S + a \rho_i^P)/(N-1)$.

If every player of the population chooses the three actions
with equal probability, namely that $\rho_i^R
= \rho_i^P = \rho_i^S = 1/3$ for $i=1, 2, \ldots, N$,
then the expected payoff for a player
is the same no matter which action she chooses in one
round of the game,
i.e., $g_i^R= g_i^P= g_i^S  = (1+a)/3$
for $i=1,2,\ldots, N$.
Then the expected payoff of a player $i$ in one game
round is $(1+a)/3$, which will not increase if the
probabilities
$\rho_i^{R}, \rho_i^{P}, \rho_i^{S}$ deviate
from $1/3$.
Therefore $\rho_i^{R}=\rho_i^{P}=\rho_i^{S}=1/3$
(for all $i=1,2,\ldots, N$) is a mixed-strategy NE
of the game.

Let us also discuss a little bit about the uniqueness
of this mixed-strategy NE.
If the payoff parameter $a \leq 1$,
this mixed-strategy NE is not unique.
We can easily check that
$\rho_i^{R}=1, \rho_i^{P}=\rho_i^{S}=0$
(for all $i=1,2,\ldots, N$) is
a pure-strategy NE. Similarly,
$\rho_i^{P}=1, \rho_i^{S}=\rho_i^{R}=0$
($i=1,2,\ldots, N$) and
$\rho_i^{S}=1, \rho_i^{R}=\rho_i^{P}=0$
($i=1,2,\ldots, N$) are two other pure-strategy
NEs.
In such a pure-strategy NE the payoff of
a player is $1$ in one game round.
This value is considerably higher than the average
payoff of $(1+a)/3$ a player will gain if
the population is in the above mentioned mixed-strategy NE.

On the other hand, if the payoff parameter $a>1$, then there is no
pure-strategy NE for the RPS game.
This is simple to prove.
Suppose the population is initially in a pure-strategy NE
with $\rho_i^{R}=1$ for $i=1,2,\ldots, N$.
If one player now shifts to action $S$, her
payoff will increase from $1$ to $a$.
Therefore this player will keep the new action $S$ in
later rounds of the game, and the original pure-strategy
NE is then destroyed.

We believe that the mixed-strategy NE of
$\rho_i^{R}=\rho_i^{P}=\rho_i^{S}=1/3$ (for $i=1,2,
\ldots, N$) is the only Nash equilibrium of our RPS game in the
 whole parameter region of $a>1$ (except for very few
 isolated values of $a$, maybe).
Unfortunately we are unable to offer a rigorous proof of
this conjecture for a generic value of population size $N$.
But this conjecture is supported by our empirical observations,
see Table~S1.

Among the $60$ experimental sessions performed at different
values of $a$, we observed that all the players in $59$
experimental sessions change their actions frequently. The
mean values of the individual action
probabilities $\rho_i^R$, $\rho_i^P$, $\rho_i^S$ are all
close to $1/3$.
(The slightly higher mean probability of choosing
action $R$ in the empirical data of Table~S1
might be linked to the fact that
``Rock" is the left-most candidate choice in each player's
decision window.)

We did notice considerable deviation from the NE mixed strategy
in one experimental session of $a=1.1$,
though. After the RPS game
has proceeded for $72$ rounds, the six players of this
exceptional session all stick to the same action R and
do not shift to the other two actions. This population
obviously has reached a highly cooperative state after $72$
game rounds with $\rho_i^R=1$ for all the six players.
As we have pointed out, such a cooperative state is not a
pure-strategy NE. We do not consider this exceptional
experimental session in the data analysis and model building
phase of this work.


\section{Evolutionary stability of the Nash equilibrium}

We now demonstrate that the mixed-strategy NE
with $\rho_i^{R}=\rho_i^{P}=\rho_i^{S}=1/3$
($i=1,2,\ldots, N$) is an evolutionarily stable strategy only
when the payoff parameter $a>2$.

To check for the evolutionary stability of this mixed-strategy
NE, let us assume a mutation occurs to the population such that
$n\geq 1$ players now adopt a mutated strategy,
while the remaining $(N-n)$ players still adopt the
NE mixed strategy. We denote by $\tilde{\rho}^{R}$
(and respectively, $\tilde{\rho}^{P}$ and
$\tilde{\rho}^{S}$) as the probability that in one round of
the game, action $R$ (respectively, $P$ and $S$)
will be chosen by a player who adopts the mutated strategy.
Obviously $\tilde{\rho}^{R}+\tilde{\rho}^{P}+\tilde{\rho}^{S}
\equiv 1$.

For a player who adopts the
NE mixed strategy, her expected payoff in one game round
is simply $g_{0}=(1+a)/3$.
On the other hand, the expected payoff $\tilde{g}$ in one
game round for a player who adopts the mutated strategy is
expressed as
$\tilde{g}  =  \tilde{\rho}^{R} \tilde{g}^{R} +
\tilde{\rho}^{P} \tilde{g}^{P}
+ \tilde{\rho}^{S} \tilde{g}^{S}$,
where $\tilde{g}^{R}$ (and respectively,
$\tilde{g}^{P}$ and $\tilde{g}^{S}$) is the expected payoff of
one play for a player in the mutated sub-population
if she chooses action $R$ (respectively, $P$ and $S$):
\begin{eqnarray}
\tilde{g}^{R} &=& \frac{N-n}{N-1} \times \frac{1+a}{3} +
\frac{n-1}{N-1} \times   \bigl(
 \tilde{\rho}^{R}+   a \tilde{\rho}^{S} \bigr) \; , \\
\tilde{g}^{P} &=& \frac{N-n}{N-1} \times \frac{1+a}{3} +
\frac{n-1}{N-1} \times  \bigl(
 \tilde{\rho}^{P}+ a \tilde{\rho}^{R} \bigr) \; , \\
\tilde{g}^{S} &=& \frac{N-n}{N-1} \times \frac{1+a}{3} +
\frac{n-1}{N-1} \times
\bigl(  \tilde{\rho}^{S}+ a \tilde{\rho}^{P} \bigr) \; .
\end{eqnarray}
Inserting these three expressions
into the expression of
$\tilde{g}$, we obtain that
\begin{eqnarray}
\tilde{g} &=&
\frac{N-n}{N-1} \times \frac{1+a}{3}
+\frac{n-1}{N-1} \times \nonumber \\
& &  \Bigl( 1 +
(a-2) \bigl[ \tilde{\rho}^{R} \tilde{\rho}^{P} +
(\tilde{\rho}^{R} + \tilde{\rho}^{P} )
(1-\tilde{\rho}^{R}-\tilde{\rho}^{P}) \bigr] \Bigr) \nonumber \\
&=& g_{0} - \frac{(a-2) (n-1)}{N-1} \times \nonumber \\
& &  \Bigl(
\bigl[ (\tilde{\rho}^{R}-1/3)+(\tilde{\rho}^{P}/2-1/6)\bigr]^2
+3 \bigl[\tilde{\rho}^{P}-1/3\bigr]^{2}/4 \Bigr) \; . \nonumber \\
& & 
\label{eq:tildeg}
\end{eqnarray}

If the payoff parameter $a>2$, we see from Eq.~[\ref{eq:tildeg}]
 that the expected payoff $\tilde{g}$ of the
mutated strategy never exceeds that of the NE mixed strategy.
Therefore the NE mixed strategy is an evolutionarily stable
strategy. Notice that the difference $(\tilde{g}-g_0)$ is proportional
to $(a-2)$, therefore the larger the value of $a$, the higher is the
cost of deviating from the NE mixed strategy.

On the other hand, in the case of $a<2$, the value of
$\tilde{g}-g_{0}$ will be positive if two or
more players adopt the mutated strategy.
Therefore the NE mixed strategy is an evolutionarily unstable
strategy.

The mixed-strategy NE for the game with payoff parameter $a=2$ is
referred to as evolutionarily neutral since it is
neither evolutionarily stable nor evolutionarily unstable.


\section{Cycling frequencies predicted by two simple models}

We now demonstrate that the empirically observed persistent
cycling behaviors could not have been observed if the
population were in the mixed-strategy NE,
and they cannot be explained by the independent decision
model either.

\subsection{Assuming the mixed-strategy Nash equilibrium}

If the population is in the mixed-strategy NE,
each player will pick an action uniformly at random
at each time $t$. Suppose the social state at time
$t$ is ${\bf s}=(n_R, n_P, n_S)$, then the probability
$M_{0}[{\bf s}^\prime | {\bf s}]$ of the
social state being ${\bf s}^\prime = (n_R^\prime, n_P^\prime,
n_S^\prime)$ at time $(t+1)$ is simply expressed as
\begin{equation}
M_0[{\bf s}^\prime | {\bf s}] =
\frac{N!}{(n_R^\prime)! (n_P^\prime)! (n_S^\prime)!} \Bigl(
\frac{1}{3}\Bigr)^{N} \; ,
\end{equation}
which is independent of ${\bf s}$. Because of this
history independence, the social state probability
distribution $P_0^{*}({\bf s})$ for
any ${\bf s}=(n_R, n_P, n_S)$ is
\begin{equation}
P_0^{*}({\bf s}) =
\frac{N!}{n_R! n_P! n_S!} \Bigl(
\frac{1}{3}\Bigr)^{N} \; .
\end{equation}

The social state
transition obeys the detailed balance condition that
$P_0^{*}({\bf s}) M_0[{\bf s}^\prime | {\bf s}]
= P_0^{*}({\bf s}^\prime ) M_0[{\bf s} | {\bf s}^\prime]$.
Therefore no persist cycling can exist in the
mixed-strategy NE. Starting from any initial social
state ${\bf s}$, the mean of the social states at
the next time step is ${\bf d}({\bf s})= \sum_{{\bf s}^\prime}
M_0[{\bf s}^\prime | {\bf s}] {\bf s}^\prime = {\bf c}_0$,
i.e., identical to the centroid of the social state plane.

\subsection{Assuming the independent decision model}

In the independent decision model, every player of the
population decides on her next action $q^\prime
 \in \{R, P, S\}$ in a probabilistic
manner based on her current action $q\in \{R, P, S\}$
only.
For example, if the current action of a player $i$ is
$R$, then in the next game round this player has probability
$R_0$ to repeat action $R$, probability $R_{-}$ to
shift action clockwise to $S$, and probability
$R_{+}$ to shift action counter-clockwise to $P$.
The transition probabilities $P_{-}, P_{0}, P_{+}$ and
$S_{-}, S_{0}, S_{+}$ are defined in the same way.
These nine transition probabilities of cause have to
satisfy the normalization conditions:
$R_{-} + R_{0} + R_{+} = 1$,
$P_{-} + P_{0} + P_{+} = 1$, and
$S_{-} + S_{0} + S_{+} = 1$.

\widetext
Given the social state ${\bf s}=(n_R, n_P, n_S)$ at time
$t$, the probability
$M_{id}[{\bf s}^\prime | {\bf s}]$ of the population's
social state being ${\bf s}^\prime = (n_R^\prime, n_P^\prime,
n_S^\prime)$ at time $(t+1)$ is
\begin{eqnarray}
M_{id}[{\bf s}^\prime | {\bf s}] &=&
\sum\limits_{n_{R\rightarrow R}}
\sum\limits_{n_{R\rightarrow P}}
\sum\limits_{n_{R\rightarrow S}}
\frac{n_R!}{n_{R\rightarrow R}!
n_{R\rightarrow P}!
n_{R\rightarrow S}! }
R_{-}^{n_{R\rightarrow S}} R_{0}^{n_{R\rightarrow R}}
R_{+}^{n_{R\rightarrow P}}
\delta_{n_{R\rightarrow R}+n_{R\rightarrow P}
+n_{R\rightarrow S}}^{n_R}
\nonumber \\
& & \times \sum\limits_{n_{P\rightarrow R}}
\sum\limits_{n_{P\rightarrow P}}
\sum\limits_{n_{P\rightarrow S}}
\frac{n_P!}{n_{P\rightarrow R}!
n_{P\rightarrow P}!
n_{P\rightarrow S}! }
P_{-}^{n_{P\rightarrow R}}
P_{0}^{n_{P\rightarrow P}}
P_{+}^{n_{P\rightarrow S}}
\delta_{n_{P\rightarrow R}+n_{P\rightarrow P}
+n_{P\rightarrow S}}^{n_P} \nonumber \\
& & \times
\sum\limits_{n_{S\rightarrow R}}
\sum\limits_{n_{S\rightarrow P}}
\sum\limits_{n_{S\rightarrow S}}
\frac{n_S!}{n_{S\rightarrow R}!
n_{S\rightarrow P}!
n_{S\rightarrow S}! }
S_{-}^{n_{S\rightarrow P}}
S_{0}^{n_{S\rightarrow S}}
S_{+}^{n_{S\rightarrow R}}
\delta_{n_{S\rightarrow R}+n_{S\rightarrow P}
+n_{S\rightarrow S}}^{n_S}
\nonumber \\
& & \times \delta^{n_R^\prime}_{n_{R\rightarrow R}
+n_{P\rightarrow R}+n_{S\rightarrow R}}
\delta^{n_P^\prime}_{n_{R\rightarrow P}
+n_{P\rightarrow P}+n_{S\rightarrow P}}
\delta^{n_S^\prime}_{n_{R\rightarrow S}
+n_{P\rightarrow S}+n_{S\rightarrow S}} \; ,
\end{eqnarray}
where $n_{q\rightarrow q^\prime}$ denotes the
total number of action transitions from $q$ to
$q^\prime$, and
 $\delta_{m}^{n}$ is the Kronecker symbol such that $\delta_{m}^{n}=1$
if $m=n$ and $\delta_{m}^{n}=0$ if $m\neq n$.

\endwidetext

For this independent decision model,
the steady-state distribution
$P_{id}^{*}({\bf s})$ of the social states is determined
by solving
\begin{equation}
\label{eq:idsteadystate}
P_{id}^{*}({\bf s}) =
\sum\limits_{{\bf s}^\prime} M_{id}[{\bf s} | {\bf s}^\prime]
P_{id}^{*}({\bf s}^\prime)
\; .
\end{equation}
When the population has reached this  steady-state
distribution,
the mean cycling frequency $f_{id}$
is then computed as
\begin{equation}
f_{id} = \sum\limits_{{\bf s}} P_{id}^{*}({\bf s})
\sum\limits_{{\bf s}^\prime} M_{id}[{\bf s}^\prime | {\bf s}]
\theta_{{\bf s}\rightarrow {\bf s}^\prime} \; ,
\end{equation}
where $\theta_{{\bf s}\rightarrow {\bf s}^\prime}$ is the
rotation angle associated with the transition
${\bf s}\rightarrow {\bf s}^\prime$, see Eq.~[7].

Using the empirically determined action
transition probabilities
of Fig.~S1 as inputs, the independent
decision model
predicts the cycling frequency to be $0.0050$ (for $a=1.1$),
$-0.0005$ ($a=2$), $-0.0024$ ($a=4$), $-0.0075$ ($a=9$)
and $-0.0081$ ($a=100$), which are all very close to zero and
significantly different from the empirical values.
Therefore the assumption of players making decisions
independently of each other
cannot explain population-level cyclic motions.


\section{Details of the conditional response model}

\subsection{Social state transition matrix}

In the most general case, our win-lose-tie conditional response (CR) model
has nine transition parameters, namely $W_{-}$, $W_{0}$, $W_{+}$,
$T_{-}$, $T_{0}$, $T_{+}$, $L_{-}$, $L_{0}$, $L_{+}$.
These parameters
are all non-negative and are constrained by three normalization
conditions:
\begin{equation}
\label{eq:wltconstraints}
W_{-}+W_{0}+W_{+} = 1 \;, 
T_{-}+T_{0}+T_{+} = 1 \; ,
L_{-}+L_{0}+L_{+} = 1 \; ,
\end{equation}
therefore the three vectors $(W_{-}, W_{0}, W_{+})$,
$(T_{-}, T_{0}, T_{+})$ and $(L_{-}, L_{0}, L_{+})$ represent
three points of the three-dimensional simplex. Because
of Eq.~[\ref{eq:wltconstraints}], we can use a set
$\Gamma \equiv \{W_{-}, W_{+}; T_{-}, T_{+}; L_{-}, L_{+}\}$ of
six transition probabilities to denote a
conditional response strategy.

The parameters $W_{+}$ and $W_{-}$ are, respectively,
the conditional probability that a player (say $i$) will
perform a counter-clockwise or clockwise action shift
in the next game round, given that she wins over the
opponent (say $j$) in the current game round.
Similarly the parameters $T_{+}$ and $T_{-}$ are the
two action shift probabilities conditional on the current
play being a tie, while $L_{+}$ and $L_{-}$ are the
action shift probabilities conditional on the
current play outcome being `lose'. The parameters
$W_{0}, T_{0}, L_{0}$ are the probabilities of a player
repeating the same action in the next play given
the current play outcome being `win', `tie' and `lose',
respectively.
For example, if the current action of $i$ is $R$ and
that of $j$ is $S$, the joint probability of $i$
choosing action $P$ and $j$ choosing action $S$ in the
next play is $W_{+} L_{0}$; while if both players choose $R$
in the current play,
the joint probability of player $i$
choosing $P$ and player $j$ choosing $S$ in the next play is
then $T_{+} T_{-}$.

We denote by ${\bf s}\equiv (n_{R}, n_{P}, n_{S})$ a social
state of the population, where $n_R$, $n_{P}$, and $n_{S}$
are the number of players who adopt action $R$, $P$ and $S$
in one round of play, respectively.
Since $n_R+n_P+n_S \equiv N$ there
are $(N+1) (N+2)/2$ such social states, all lying
on a three-dimensional plane bounded by an
equilateral triangle  (Fig.~1C).

Furthermore we denote by $n_{r r}$, $n_{p p}$,
$n_{s s}$, $n_{r p}$, $n_{p s}$ and $n_{s r}$, respectively,
the number of pairs in which the competition being $R$--$R$,
$P$--$P$,  $S$--$S$, $R$--$P$, $P$--$S$, and $S$--$R$, in this
round of play.
These nine integer values are not independent but are
related by the following equations:
\begin{eqnarray}
n_R & = & 2 n_{r r} + n_{s r} + n_{r p} \; , \nonumber \\
n_{P} & = & 2 n_{p p} + n_{r p} + n_{p s} \; , \nonumber \\
n_{S} & = & 2 n_{s s} + n_{p s} + n_{s r} \; . \nonumber \\
& & 
\label{eq:nRPS}
\end{eqnarray}

Knowing the values of $n_{R}, n_{P}, n_{S}$
is not sufficient to uniquely fix the values of
$n_{r r}, n_{p p}, \ldots, n_{s r}$.
the conditional joint
probability distribution of $n_{r r}, n_{p p}, n_{s s}, n_{r p},
n_{p s}, n_{s r}$ is expressed as Eq.~[3].
To understand this expression, let us first notice that
the total number of pairing patterns of $N$ players is equal to
$$
\frac{ N! }{ (N/2)! \; 2^{N/2}} = (N-1)!! \; ,
$$
which is independent of the specific values of $n_{R}, n_{P}, n_{S}$; 
and second, the number of pairing patterns with
$n_{r r}$ $R$--$R$ pairs, $n_{p p}$ $P$--$P$ pairs, $\ldots$,
and $n_{s r}$ $S$--$R$ pairs is equal to
$$
\frac{ n_{R}! \; n_{P}!\; n_{S}!}{
 2^{n_{r r}} n_{r r}! \; 2^{n_{p p}} n_{p p}! \;
2^{n_{s s}}  n_{s s}! \; n_{r p}! \; n_{p s}! \; n_{s r}!} \; .
$$

Given the values of $n_{r r}, n_{p p}, \ldots, n_{s r}$ which
describe the current pairing pattern, the conditional probability of the
social state in the next round of play can be determined. We
just need to carefully analyze the conditional probability for
each player of the population.
For example, consider a $R$--$P$ pair at game round $t$.
This is a lose--win pair, therefore the two involved players
will determine their actions of the next game
round according to the CR parameters $(L_-, L_0, L_+)$
and  $(W_-, W_0, W_+)$, respectively.
At time $(t+1)$ there are six possible outcomes:
(rr) both players take action $R$, with probability $L_0 W_{-}$;
 (pp) both players take action $P$, with probability
$L_{+} W_{0}$;
(ss) both players take action $S$, with probability
$L_{-}W_{+}$;
(rp) one player takes action $R$ while the other takes
action $P$, with probability
$(L_0 W_0 + L_{+} W_{-})$;
(ps) one player takes action $P$ while the other takes
action $S$, with probability
$(L_{+} W_{+} + L_{-} W_{0})$;
(sr) one player takes action $S$ and the other takes action $R$,
with probability $(L_{-} W_{-} + L_{0} W_{+})$.
Among the $n_{rp}$
$R$--$P$ pairs of time $t$, let us assume that after the
play, $n_{rp}^{rr}$ of these pairs will outcome  (rr),
$n_{rp}^{pp}$ of them will outcome  (pp),
$n_{rp}^{ss}$ of them will outcome (ss),
$n_{rp}^{rp}$ of them will outcome (rp),
$n_{ps}^{ps}$ of them will outcome (ps),
and $n_{sr}^{sr}$ of them will outcome (sr). Similarly we
can define a set of non-negative integers to describe the
outcome pattern for each of the other five types of pairs.

\widetext

Under the random pairwise-matching game
protocol, our conditional response model gives the following
expression for the transition probability
$M_{cr}[{\bf s}^\prime  | {\bf s}]$
from the social state ${\bf s} \equiv (n_{R}, n_{P}, n_{S})$ at
time $t$ to the social state ${\bf s}^\prime \equiv
(n_{R}^\prime, n_{P}^\prime, n_{S}^\prime)$ at time $(t+1)$:
\begin{eqnarray}
& & \hspace*{-1.0cm}
M_{cr}[{\bf s}^\prime  | {\bf s}] = \nonumber \\
& & \sum\limits_{n_{rr}, n_{pp},\ldots, n_{sr}}
\frac{n_{R}!\; n_{P}!\; n_{S}! \;
\delta_{2 n_{rr} + n_{sr} + n_{rp}}^{n_{R}}
\; \delta_{2 n_{pp} + n_{rp} + n_{ps}}^{n_{P}}
\; \delta_{2 n_{ss} + n_{ps} + n_{sr}}^{n_{S}}}{ (N-1)!! \;
 2^{n_{rr}} n_{rr}! \; 2^{n_{pp}} n_{pp}! \;
 2^{n_{ss}} n_{ss}! \; n_{rp}! \; n_{ps}! \; n_{sr}!}   \nonumber \\
& & \times
\sum\limits_{n_{rr}^{rr}, \ldots, n_{rr}^{sr}}
\frac{ n_{rr}!\; T_0^{2 n_{rr}^{rr}}\;
T_+^{2 n_{rr}^{pp}} \;
T_-^{2 n_{rr}^{ss}} \;
 (2 T_+ T_0)^{n_{rr}^{rp}}\;
(2 T_+ T_-)^{n_{rr}^{ps}} \; (2 T_0 T_-)^{n_{rr}^{sr}}
}{n_{rr}^{rr} ! \; n_{rr}^{pp} ! \; n_{rr}^{ss} ! \;
n_{rr}^{rp} ! \; n_{rr}^{ps}! \; n_{rr}^{sr}!}
\delta_{n_{rr}^{rr} + \ldots + n_{rr}^{sr}}^{n_{rr}} \nonumber \\
& & \times
\sum\limits_{n_{pp}^{rr}, \ldots, n_{pp}^{sr}}
\frac{ n_{pp}!\; T_-^{2 n_{pp}^{rr}}\;
T_0^{2 n_{pp}^{pp}} \;
T_+^{2 n_{pp}^{ss}} \;
 (2 T_0 T_-)^{n_{pp}^{rp}}\;
(2 T_+ T_0)^{n_{pp}^{ps}} \; (2 T_+ T_-)^{n_{pp}^{sr}}
}{n_{pp}^{rr} ! \; n_{pp}^{pp} ! \; n_{pp}^{ss} ! \;
n_{pp}^{rp} ! \; n_{pp}^{ps}! \; n_{pp}^{sr}!}
\delta_{n_{pp}^{rr} + \ldots + n_{pp}^{sr}}^{n_{pp}} \nonumber \\
 & & \times
\sum\limits_{n_{ss}^{rr}, \ldots, n_{ss}^{sr}}
\frac{ n_{ss}!\; T_+^{2 n_{ss}^{rr}}\;
T_-^{2 n_{ss}^{pp}} \;
T_0^{2 n_{ss}^{ss}} \;
 (2 T_+ T_-)^{n_{ss}^{rp}}\;
(2 T_0 T_-)^{n_{ss}^{ps}} \; (2 T_+ T_0)^{n_{ss}^{sr}}
}{n_{ss}^{rr} ! \; n_{ss}^{pp} ! \; n_{ss}^{ss} ! \;
n_{ss}^{rp} ! \; n_{ss}^{ps}! \; n_{ss}^{sr}!}
\delta_{n_{ss}^{rr} + \ldots + n_{ss}^{sr}}^{n_{ss}} \nonumber \\
& & \times
\sum\limits_{n_{rp}^{rr}, \ldots, n_{rp}^{sr}}
\frac{ n_{rp}! \;
\delta_{n_{rp}^{rr} + \ldots + n_{rp}^{sr}}^{n_{rp}} }
{n_{rp}^{rr} ! \; n_{rp}^{pp} ! \; n_{rp}^{ss}! \;
n_{rp}^{rp} ! \; n_{rp}^{ps}! \; n_{rp}^{sr}!}
(W_{-} L_{0} )^{n_{rp}^{rr}}\;
(W_{0} L_{+})^{n_{rp}^{pp}} \;
(W_{+} L_{-})^{n_{rp}^{ss}} \;
\nonumber \\
& & \quad\quad \quad\quad \quad\quad \times \;
 (W_0 L_0 + W_{-} L_{+})^{n_{rp}^{rp}}\;
(W_{+} L_{+} + W_{0} L_{-})^{n_{rp}^{ps}} \;
(W_{+} L_{0} + W_{-} L_{-})^{n_{rp}^{sr}}
 \nonumber \\
& & \times
\sum\limits_{n_{ps}^{rr}, \ldots, n_{ps}^{sr}}
\frac{ n_{ps}!\; \delta_{n_{ps}^{rr} + \ldots
+ n_{ps}^{sr}}^{n_{ps}} }
{n_{ps}^{rr} ! \; n_{ps}^{pp} ! \; n_{ps}^{ss} ! \;
n_{ps}^{rp} ! \; n_{ps}^{ps}! \; n_{ps}^{sr}!}
(W_{+} L_{-})^{n_{ps}^{rr}}\;
(W_- L_0)^{n_{ps}^{pp}} \;
(W_0 L_+)^{n_{ps}^{ss}} \nonumber \\
& & \quad\quad \quad\quad \quad\quad \times \;
 (W_+ L_0 + W_- L_-)^{n_{ps}^{rp}}\;
(W_0 L_0 + W_- L_+)^{n_{ps}^{ps}} \;
(W_+ L_+  +W_0 L_- )^{n_{ps}^{sr}}
 \nonumber \\
& & \times
\sum\limits_{n_{sr}^{rr}, \ldots, n_{sr}^{sr}}
\frac{ n_{sr}!\; \delta_{n_{sr}^{rr} + \ldots + n_{sr}^{sr}}^{n_{ps}} }
{n_{sr}^{rr} ! \; n_{sr}^{pp} ! \; n_{sr}^{ss} ! \;
n_{sr}^{rp} ! \; n_{sr}^{ps}! \; n_{sr}^{sr}!}
(W_0 L_+)^{n_{sr}^{rr}}\;
(W_+ L_-)^{n_{sr}^{pp}} \;
(W_- L_0)^{n_{sr}^{ss}} \nonumber \\
& & \quad\quad \quad\quad \quad\quad \times\;
 (W_+ L_+ + W_0  L_-)^{n_{sr}^{rp}}\;
(W_+ L_0 + W_- L_-)^{n_{sr}^{ps}} \;
(W_0 L_0 +W_- L_+)^{n_{sr}^{sr}}
 \nonumber \\
 & & \times \; \delta^{n_{R}^\prime}_{2(n_{rr}^{rr}+n_{pp}^{rr}+n_{ss}^{rr}+
n_{rp}^{rr}+n_{ps}^{rr}+n_{sr}^{rr})
+(n_{rr}^{sr}+n_{pp}^{sr}+n_{ss}^{sr}+n_{rp}^{sr}
+n_{ps}^{sr}+n_{sr}^{sr})
+(n_{rr}^{rp}+n_{pp}^{rp}+n_{ss}^{rp}+n_{rp}^{rp}+n_{ps}^{rp}
+n_{sr}^{rp})} \nonumber \\
& & \times \; \delta^{n_{P}^\prime}_{2(n_{rr}^{pp}+n_{pp}^{pp}+n_{ss}^{pp}+
n_{rp}^{pp}+n_{ps}^{pp}+n_{sr}^{pp})
+(n_{rr}^{rp}+n_{pp}^{rp}+n_{ss}^{rp}+n_{rp}^{rp}+n_{ps}^{rp}
+n_{sr}^{rp})+(n_{rr}^{ps}+n_{pp}^{ps}+n_{ss}^{ps}+n_{rp}^{ps}
+n_{ps}^{ps}+n_{sr}^{ps})} \nonumber \\
& & \times \; \delta^{n_{S}^\prime}_{2(n_{rr}^{ss}
+n_{pp}^{ss}+n_{ss}^{ss}+
n_{rp}^{ss}+n_{ps}^{ss}+n_{sr}^{ss})
+(n_{rr}^{ps}+n_{pp}^{ps}+n_{ss}^{ps}+n_{rp}^{ps}+n_{ps}^{ps}
+n_{sr}^{ps})+(n_{rr}^{sr}+n_{pp}^{sr}+n_{ss}^{sr}+n_{rp}^{sr}
+n_{ps}^{sr}+n_{sr}^{sr})} \; .
\nonumber \\
& &
\label{eq:Transition}
\end{eqnarray}

\subsection*{Steady-state properties}

It is not easy to further simplify the
transition probabilities $M_{cr}[{\bf s}^\prime | {\bf s}]$, but
their values can be determined numerically.
Then the steady-state distribution
$P_{cr}^{*}({\bf s})$ of the social states is determined by numerically
solving the following equation:
\begin{equation}
\label{eq:steadystate}
P_{cr}^{*}({\bf s}) =
\sum\limits_{{\bf s}^\prime} M_{cr}[{\bf s} | {\bf s}^\prime]
P_{cr}^{*}({\bf s}^\prime)
\; .
\end{equation}
Except for extremely rare cases of the conditional
response parameters (e.g.,
$W_{0}=T_{0}=L_{0}=1$), the Markov transition
matrix defined by Eq.~[\ref{eq:Transition}] is
ergodic, meaning that it is possible to reach from
any social state ${\bf s}_1$ to any another
social state ${\bf s}_2$ within a finite number of time
steps.
This ergodic property guarantees
that Eq.~[\ref{eq:steadystate}] has
a unique  steady-state solution $P^{*}({\bf s})$.
In the steady-state,
the mean cycling frequency $f_{cr}$ of this conditional response
model is then computed through Eq.~[4] of the main text.
And the mean payoff $g_{cr}$ of each player in one game round
is obtained by
\begin{eqnarray}
g_{cr}  & = & \frac{1}{N} \sum\limits_{{\bf s}}
P_{cr}^{*}({\bf s}) \sum\limits_{n_{r r}, n_{p p},
 \ldots , n_{r s}} {\rm Prob}_{{\bf s}}(n_{r r}, n_{p p},
 \ldots, n_{s r}) 
 \bigl[
2 (n_{r r}+ n_{p p} + n_{s s}) + a (n_{r p} + n_{p s} +
n_{s r}) \bigr]  \nonumber \\
& = & 1 +
\frac{(a-2)}{N}
\sum\limits_{{\bf s}}
P_{cr}^{*}({\bf s}) \sum\limits_{n_{r r}, n_{p p},
 \ldots , n_{r s}} {\rm Prob}_{{\bf s}}(n_{r r}, n_{p p},
 \ldots, n_{s r})  [n_{r p} + n_{p s} + n_{s r}] \; . \label{eq:gcr2}
\end{eqnarray}
The expression [\ref{eq:gcr2}] is identical to Eq.~[5] of the main text.
\endwidetext

Using the five sets of CR parameters of Fig.~2,
we obtain the values of $g_{cr}$ for
the five data sets to be
$g_{cr}=g_{0}+0.005$ (for $a=1.1$),
$g_{cr}=g_0$ ($a=2$),
$g_{cr}=g_0+0.001$ ($a=4$),
$g_{cr}=g_0+0.004$ ($a=9$),
and $g_{cr}=g_0+0.08$ ($a=100$).
When $a\neq 2$ the predicted values of $g_{cr}$ are
all slightly higher than $g_0=(1+a)/3$, which is the
expected payoff per game round for a player adopting the
NE mixed strategy.
On the empirical side, we compute the mean payoff $g_i$ per
game round for each player $i$ in all populations of the
same value of $a$. The mean value of $g_i$ among these
players, denoted as $\overline{g}$, is also found to be
slightly
higher than $g_0$ for all the four sets of populations of
$a\neq 2$. To be more specific, we observe that
$\overline{g}-g_0$ equals to
$0.009 \pm 0.004$ (for $a=1.1$, mean
$\pm$ SEM), $0.000\pm 0.006$ ($a=2$),
$0.004 \pm 0.012$ ($a=4$), $0.01 \pm 0.02$ ($a=9$)
and $0.05 \pm 0.37$ ($a=100$).
These theoretical and empirical results indicate that
the conditional response strategy has the potential of
bringing higher payoffs to individual players as compared
with the NE mixed strategy.

\subsection{The symmetric case}

Very surprisingly, we find that asymmetry in the CR parameters
is not essential for cycle persistence and direction.
We find that if the CR parameters are symmetric with
respect to clockwise and counter-clockwise action shifts
(namely, $W_{+}/W_{-} = T_{+} / T_{-} = L_{+} / L_{-} =1$),
the cycling frequency $f_{cr}$ is still nonzero as
long as $W_{0} \neq L_0$. The magnitude of $f_{cr}$
increases with $|W_{0}-L_{0}|$ and decreases with
$T_0$, and the cycling  is counter-clockwise ($f_{cr}>0$)
if $W_0 > L_0$ and clockwise ($f_{cr}<0$) if $L_0 > W_0$, see
Fig.~S4 C.
In other words, in this symmetric CR model,
if losers are more (less) likely to shift actions than winners, the social
state cycling will be counter-clockwise (clockwise).

To give some concrete examples, we symmetrize the transition
parameters of Fig.~2 (F--J) while
keeping the empirical values of
$W_{0}, T_{0}, L_{0}$ unchanged. The resulting
cycling frequencies are, respectively,
$f_{cr}= 0.024$ ($a=1.1$), $0.017$ ($a=2.0$),
$0.017$ ($a=4.0$), $0.015$ ($a=9.0$) and $0.017$ ($a=100.0$), which are all
significantly beyond zero.
Our model is indeed
dramatically different from the best response model, for which
asymmetry in decision-making is a basic assumption.

\subsection{Sampling the conditional response parameters}

For the population size $N=6$, we uniformly sample
$2.4 \times 10^9$ sets of conditional response parameters
$W_{-}, W_{0}, \ldots, L_{0}, L_{+}$ under the constraints of
Eq.~[\ref{eq:wltconstraints}], and for each of them we
determine the theoretical frequency $f_{cr}$ and the
theoretical payoff $g_{cr}$ numerically.
By this way we
obtain the joint probability distribution of $f_{cr}$ and
$g_{cr}$ and also the marginal probability distributons of
$f_{cr}$ and $g_{cr}$, see Fig.~3,  Fig.~S2 and Fig.~S4 A. 
The mean values of $|f_{cr}|$ and $g_{cr}$ are
then computed from this joint probability distribution.
We find that the mean value of $f_{cr}$ is equal to zero, while
the mean value of $|f_{cr}| \approx 0.061$.

The mean value of $g_{cr}$ for randomly sampled CR strategies
is determined to be $g_0 - 0.0085 (a-2)$ for $N=6$.
When $a>2$ this mean value is
less than $g_0$,
indicating that if the CR parameters are randomly chosen,
the CR strategy has high probability of
being inferior to the NE mixed strategy.

However, we also notice that $g_{cr}$ can considerably
exceed $g_0$ for some optimized sets of conditional
response parameters (see Fig.~3 for the case of $N=6$ and
Fig.~S2 for the case of $N=12$). To give some
concrete examples, here we list for population size $N=6$
the five sets of CR parameters
of the highest values of $g_{cr}$ among the sampled $2.4 \times 10^9$
sets of parameters:
\begin{enumerate}
\item[1.] $\{W_{-}=0.002, W_{0}=0.998, W_{+}=0.000,
T_{-}=0.067, T_{0}=0.823, T_{+}=0.110,
L_{-}=0.003, L_{0}=0.994, L_{+}=0.003\}$. For
this set, the cycling frequency is $f_{cr}=0.003$, and the
expected payoff of one game round is
$g_{cr}=g_0+0.035  (a-2)$.

\item[2.]$\{W_{-}=0.001, W_{0}=0.993, W_{+}=0.006,
T_{-}=0.154, T_{0}=0.798, T_{+}=0.048,
L_{-}=0.003, L_{0}=0.994, L_{+}=0.003\}$. For
this set, $f_{cr}=0.007$ and $g_{cr}=g_0+0.034 (a-2)$.

\item[3.] $\{W_{-}=0.995, W_{0}=0.004, W_{+}=0.001,
T_{-}=0.800, T_{0}=0.142, T_{+}=0.058,
L_{-}=0.988, L_{0}=0.000, L_{+}=0.012\}$. For this set,
$f_{cr}=-0.190$ and
$g_{cr}=g_0 + 0.034 (a-2)$.

\item[4.] $\{W_{-}=0.001, W_{0}=0.994, W_{+}=0.004,
T_{-}=0.063, T_{0}=0.146, T_{+}=0.791,
L_{-}=0.989, L_{0}=0.010, L_{+}=0.001\}$.
For this set, $f_{cr}=0.189$ and $g_{cr}=g_0+0.033
(a-2)$.

\item[5.] $\{W_{-}=0.001, W_{0}=0.992, W_{+}=0.006,
T_{-}=0.167, T_{0}=0.080, T_{+}=0.753, L_{-}=0.998,
L_{0}=0.000, L_{+}=0.002\}$. For this set,
$f_{cr}=0.179$ and $g_{cr}=g_0+0.033 (a-2)$.
\end{enumerate}

To determine the influence of each of the nine conditional
response parameters to the cycling frequency $f_{cr}$,
we fix each of these nine conditional
response parameters and sample all the others uniformly at
random under the constraints of
Eq.~[\ref{eq:wltconstraints}]. The mean value
$\langle f_{cr} \rangle$ of $f_{cr}$
as a function of this fixed conditional response parameter
is then obtained by repeating this process many times, see Fig.~S4 B.
As expected, we find that when the fixed conditional response
parameter is equal to $1/3$, the mean cycling frequency
$\langle f_{cr} \rangle =0$. Furthermore we find that
\begin{enumerate}
\item[1.]
If $W_{0}$, $T_{+}$ or $L_{+}$ is the fixed parameter, then
$\langle f_{cr} \rangle$  increases (almost linearly)
with fixed parameter, indicating that a larger value of
$W_{0}$, $T_{+}$ or $L_{+}$ promotes counter-clockwise cycling
at the population level.

\item[2.]
 If $W_{-}$, $T_{-}$ or $L_{0}$ is the fixed parameter,
 then $\langle f_{cr} \rangle$ decreases (almost linearly)
 with this fixed parameter, indicating that a larger value of
$W_{-}$, $T_{-}$ or $L_{0}$ promotes clockwise cycling
at the population level.

\item[3.]
If $W_{+}$, $T_{0}$ or $L_{-}$ is the
fixed parameter, then $\langle f_{cr} \rangle$ does not
change with this fixed parameter (i.e.,
$\langle f_{cr} \rangle = 0$), indicating that these
three conditional response parameters are neutral as the
cycling direction is concerned.
\end{enumerate}

\subsection{Action marginal distribution of a single player}

The social state transition matrix Eq.~[\ref{eq:Transition}] has the
following rotation symmetry:
\begin{eqnarray}
\label{eq:symmetry}
& & M_{cr}[(n_R^\prime, n_P^\prime, n_S^\prime) |
  (n_R, n_P, n_S) ] \nonumber \\
& & \quad =  M_{cr}[(n_S^\prime, n_R^\prime, n_P^\prime) |
  (n_S, n_R, n_P)] \nonumber \\
&   & \quad =  M_{cr}[(n_P^\prime, n_S^\prime, n_R^\prime) |
  (n_P, n_S, n_R)] \; .
\end{eqnarray}
Because of
this rotation symmetry, the steady-state
distribution $P_{cr}^{*}({\bf s})$ has also the rotation
symmetry that
\begin{equation}
\label{eq:psymm}
P_{cr}^{*}(n_R, n_P, n_S) = P_{cr}^{*}(n_S, n_R, n_P)
= P_{cr}^{*}(n_P, n_S, n_R) \; .
\end{equation}

After the social states of the population has reached the
steady-state distribution $P_{cr}^{*}({\bf s})$,
the probability $\rho_{cr}^{R}$
that a randomly chosen player adopts action
$R$
in one game round is expressed as
\begin{equation}
\rho_{cr}^{R} = \sum\limits_{{\bf s}} P_{cr}^{*}({\bf s}) \frac{n_R}{
n_R + n_P+ n_S}
= \frac{1}{N} \sum\limits_{{\bf s}} P_{cr}^{*}({\bf s}) n_{R}
\; ,
\end{equation}
where the summation is over all the possible social states
${\bf s}=(n_R, n_P, n_S)$. The probabilities $\rho_{cr}^P$ and
$\rho_{cr}^S$ that a randomly chosen player adopts action
$P$ and $S$ in one play can be computed similarly.
Because of the rotation symmetry
Eq.~[\ref{eq:psymm}] of $P_{cr}^{*}({\bf s})$, we obtain that
$\rho_{cr}^{R}=\rho_{cr}^{P}=\rho_{cr}^{S}=1/3$.

Therefore, if the players of the population all play the
same CR strategy, then after the population reaches the
the steady-state, the action marginal distribution of
each player will be identical to the NE mixed strategy.
In other words, the CR strategy can not be distinguished
from the NE mixed strategy through measure the action
marginal distributions of individual players.

\subsection{Computer simulations}

All of the theoretical predictions of the CR model 
have been confirmed by extensive
computer simulations.
In each of our computer simulation processes,
a population of $N$ players
repeatedly play the RPS game under the random pairwise-matching
protocol. At each game round, each player of this population
makes a choice on her action following exactly the
CR strategy. The parameters
$\{W_{-}, W_{0}, \ldots, L_{+}\}$ of this strategy is specified
at the start of the simulation and they do not change during the
simulation process.


\section{The generalized conditional response model}

If the payoff matrix of the RPS model
is more complex than the one
shown in Fig.~1A, the conditional response model may still be
applicable after some appropriate extensions.
In the most general case, we
can assume that a player's decision is influenced by the
player's own action and the opponent's action in the
 previous game round.

Let us denote by $q_{s} \in \{R, P, S\}$ a player's action
 at time $t$, and by
 $q_{o} \in \{R, P, S\}$ the action of this player's opponent
at time $t$.
Then at time $(t+1)$,
the probability that this player adopts action
$q \in \{R, P, S\}$ is denoted as
$Q_{(q_{s}, q_{o})}^{q}$, with the normalization
condition that
\begin{equation}
  \label{eq:gnorm}
  Q_{(q_{s}, q_{o})}^{R}+
  Q_{(q_{s}, q_{o})}^{P}+
  Q_{(q_{s}, q_{o})}^{S} \equiv 1 \; .
\end{equation}

This generalized conditional response model has
$27$ transition parameters, which are constrained
by $9$ normalization conditions (see Eq.~[\ref{eq:gnorm}]).
The social state transition matrix of this generalized
model is slightly more complicated than
Eq.~[\ref{eq:Transition}].

The win-lose-tie conditional response model is a limiting
case of this more general model. It can be
derived from this general model by assuming
$Q_{(R, R)}^{q}=Q_{(P, P)}^{q}=Q_{(S, S)}^{q}$,
$Q_{(R, P)}^{q}=Q_{(P, S)}^{q}=Q_{(S, R)}^{q}$,
and
$Q_{(R, S)}^{q}=Q_{(P, R)}^{q}=Q_{(S, P)}^{q}$.
These additional assumptions are reasonable only
for the simplest payoff matrix shown in Fig.~1A.

\end{document}